\documentclass[prd,aps,a4paper,twocolumn,nofootinbib]{revtex4-1}

\usepackage{graphicx,psfrag}
\usepackage{mathrsfs}
\usepackage{amsmath,amsfonts,amssymb}
\usepackage{multirow}
\usepackage{comment}
\usepackage{bm}
\usepackage{graphicx}
\usepackage{psfrag}
\usepackage{epstopdf}
\usepackage{color}
\usepackage{mathrsfs}
\usepackage{ulem}
\usepackage{hyperref}
\usepackage{diagbox} 
\usepackage{amsmath,amssymb,amsthm,wasysym}
\usepackage{capt-of}
\usepackage[utf8]{inputenc}

\def\ph{\phi}

\def\hP{{\hat{P}}}

\def\O{\mathcal{O}}

\newcommand{\be}{\begin{equation}}
\newcommand{\ee}{\end{equation}}

\newcommand{\RN}[1]{%
  \textup{\uppercase\expandafter{\romannumeral#1}}%
}

\definecolor{gray}{rgb}{0.5,0.5,0.5}
\definecolor{cyan}{rgb}{0,0.9,0.9}
\definecolor{orange}{rgb}{0.9,0.5,0}
\definecolor{magenta}{rgb}{1,0,1}
\definecolor{purple}{rgb}{0.8,0.4,0.8}
\definecolor{darkgreen}{rgb}{0,.6,0}
\definecolor{turquoise}{rgb}{0.25,0.88,0.82}

\newcommand{\RM}[1]{\MakeUppercase{\romannumeral #1{}}}

\begin{document}

\interfootnotelinepenalty=10000
\raggedbottom

\title{Spinning test-body orbiting around Schwarzschild black hole: \\
  circular dynamics and gravitational-wave fluxes}

\author{Enno Harms${}^1$, Georgios Lukes-Gerakopoulos${}^2$, Sebastiano Bernuzzi${}^3$, Alessandro Nagar${}^4$}
\affiliation{${}^1$Theoretical Physics Institute, University of Jena, 07743 Jena, Germany}
\affiliation{${}^2$Institute of Theoretical Physics, Faculty of Mathematics and Physics, Charles University in Prague, 18000 Prague, Czech Republic}
\affiliation{${}^3$DiFeST, University of Parma, and INFN, 43124, Parma, Italy}
\affiliation{${}^4$Institut des Hautes Etudes Scientifiques, 91440 Bures-sur-Yvette, France}

\begin{abstract}
We consider a spinning test-body in circular motion around a
nonrotating black hole and analyze different
prescriptions for the body's dynamics.
We compare, for the first time, the Mathisson-Papapetrou
formalism under the Tulczyjew spin-supplementary-condition (SSC), the Pirani SSC
and the Ohashi-Kyrian-Semerak SSC, and the spinning particle limit of the
effective-one-body Hamiltonian of [Phys.~Rev.~D.90,~044018(2014)]. 
We analyze the four different dynamics in terms of the ISCO shifts
and in terms of the coordinate invariant binding energies, separating
higher-order spin contributions from spin-orbit contributions. 
The asymptotic gravitational wave fluxes produced by the spinning body are computed 
by solving the inhomogeneous $(2+1)D$
Teukolsky equation and contrasted for the different cases. 
For small orbital frequencies $\Omega$, all the prescriptions
reduce to the same dynamics and the same radiation fluxes. 
For large frequencies, ${x \equiv (M \Omega)^{2/3} >0.1 }$, where $M$ is the black hole mass,
and especially for positive spins (aligned with orbital angular momentum)
a significant disagreement between the different dynamics is observed. 
The ISCO shifts can differ up to a factor two for large positive
spins; for the Ohashi-Kyrian-Semerak and the Pirani SSC the ISCO diverges around
dimensionless spins $\sim0.52$ and $\sim0.94$ respectively.
In the spin-orbit part of the energetics the deviation from the
Hamiltonian dynamics is largest for the Ohashi-Kyrian-Semerak SSC; 
it exceeds $10\%$ for $x>0.17$. The Tulczyjew and the Pirani SSCs behave compatible
across almost the whole spin and frequency range.
Our results will have direct application in including spin effects to
effective-one-body waveform models for circularized binaries in the extreme-mass-ratio limit.
\end{abstract}


\maketitle

\section{Introduction}
\label{sec:intro}

The motion of a small, spinning test-body on a fixed background
is a long-standing problem
in general relativity~\cite{Mathisson:1937zz,Papapetrou:1951pa}.
One starts with the idea to represent the motion of the small body
by the worldline of a single reference point that lies inside the body,
thus motivating the term ``spinning point-particle''.
To account for finite size effects like the \textit{spin},
one usually endows the particle with Mathisson's~\cite{Mathisson:1937zz,Mathisson:2010,tulczyjew1959motion}
``gravitational skeleton''; a multipole expansion of the energy-momentum tensor
at the reference point that sustains the appearing multipole moments up to some order. 
In the widely used pole-dipole approximation one truncates this expansion
at first order~\cite{Mathisson:2010, Steinhoff:2010zz},
neglecting quadrupolar and higher
moments~\cite{Hinderer:2013uwa, Bini:2008zzf, Bini:2013uwa, Bini:2014wua, Bini:2014epa}.
The zeroth multipole moments, often called the mass-monopole,
can be encoded in the four-momenta $p_\mu$, while
the first moments, often called the spin-dipole,
in the antisymmetric spin-tensor $S^{\mu\nu}$.
Thus the evolution system for a spinning particle
typically comprises the variables
\begin{align}
 \label{eq:spinning_particle_vars}
  \{X^\mu,v^\mu,p_\mu,S^{\mu\nu} \} \; ,
\end{align}
where $X^\mu=X^\mu(\lambda)$ is the worldline of the particle
with $\lambda$ the proper time and $v^\mu=dX^\mu/d\lambda$
the tangent vector.
The equations-of-motion (EOM) for this set of variables
can be derived from the covariant conservation of energy and momentum,
and they are called the ``Mathisson-Papapetrou-Dixon''-Equations
(MP)~\cite{Papapetrou:1951pa,tulczyjew1959motion,Dixon:1964, Dixon:1970zza,Dixon:1970zz,dixon1974dynamics,Wald:1972sz}.

The MP are not a closed system of equations with respect to the variables~\eqref{eq:spinning_particle_vars};
they prescribe the evolution of $p_\mu$ and $S^{\mu\nu}$, but not of $v^\mu$.
Thus, the EOM of a spinning particle are \textit{not} unique.
The physical reason is that there is a freedom in choosing the particle's reference point
due to the absence of a unique center-of-mass in general relativity.
To remove this ambiguity,
one might naively identify the particle with any point inside the body and then directly prescribe its tangent vector at all times.
Such an \textit{ad hoc} procedure would, however, be prone to undesired features like a worldline
that moves through the body uncontrolledly.
Instead, a physically robust procedure is to identify the reference point for the particle
with the center-of-mass as perceived by some preferred observer.
This point is called ``the centroid'', 
and it conventionally serves not only as the reference point for the particle
but also as the central point against which the internal rotations of the body, i.e.\ the spin, are measured.
Hence, selecting a centroid also fixes the particle and its spin.
The just described procedure is realized by enhancing the MP 
by a \textit{spin-supplementary condition} (SSC).
In general, a SSC imposes that $S^{0\mu}=0$ holds for some preferred observer.
Though being at first sight only a condition on the spin, it turns out that $S^{0\mu}=0$ in fact
guarantees that the observer's centroid 
is identified with the particle and used for measuring the spin~\cite{Kyrian:2007zz}.
The remaining ambiguity lies thus in the choice of the preferred observer.

Several such SSCs have been used in the
literature~\cite{Corinaldesi:1951pb,Newton:1949cq,Pirani:1956tn,Dixon:1964,tulczyjew1959motion,Ohashi:2003,Kyrian:2007zz}
and their influences on the dynamics have been studied in,
e.g.,~\cite{Kyrian:2007zz,Lukes-Gerakopoulos:2014dma}.
The variety of EOM for a spinning particle is even larger than the variety of SSCs 
because alternatively to the MP one may take a Hamiltonian approach.
Hamiltonian dynamics for a spinning particle were derived
in several different forms~\cite{Barausse:2009aa,Barausse:2009xi,Ramirez:2014,Kunst:2016tla,Bini:2015xua,Vines:2016unv}.
The mutual relations between the different dynamical approaches are presently not a trivial issue.
In~\cite{Barausse:2009aa} the theoretical equivalence of the (unclosed) MP and the Hamiltonian EOM was shown.
However, in practice one has to work with the closed MP,
i.e.\ with some fixed SSC,
and one may make a choice other than the Newton-Wigner SSC,
which most of the explicit functions of the Hamiltonian formulation have been derived for
\cite{Barausse:2009aa,Barausse:2009xi,Kunst:2016tla,Vines:2016unv}.
Even a numerical comparison of the different dynamical
prescriptions is difficult. On the one hand, conclusive comparisons require
initial data that corresponds to the same physical situation, which
is a highly non-trivial task, e.g.\ due to the shifts of the different centroids
associated with the different SSCs~\cite{Kyrian:2007zz}.
On the other hand, even when equivalent initial data are found, the respective worldlines of the
particle will sooner or later deviate from one another, thus preventing a
consistent mutual comparison~\cite{Lukes-Gerakopoulos:2014dma,Kyrian:2007zz}.
Furthermore, it has been found numerically that the dynamics of different
approaches are compatible for small spins but can diverge for large spins~\cite{Lukes-Gerakopoulos:2014dma}.
In fact, it is impossible to say that one or the other dynamical prescription
is more appropriate than the others.
But one should put forward the questions how 
the different formulations relate to each other
and whether they lead to different gravitational wave (GW) signals
when applied to the same physical situation.

In this paper we specifically consider the situation of a spinning
particle moving on a circular orbit in a nonrotating black hole (BH) background.
Such a system can be viewed as a model for a circularized spinning BH
binary of extreme-mass-ratio in which the test body is a {\it test black hole}.
Note though that at the pole-dipole level calling the body a black hole is just
a matter of view: without accounting for higher multipoles the structure of the
object is irrelevant for the dynamics and a black hole can not be discerned from
any other spinning object. In fact, one may expect the higher multipoles to become
important in the strong-field, where at least the effects from the quadrupole
term may become comparable with the effects from the dipole term. Thus, the
absence of the higher multipoles can be a reason for the discrepancies between
the different prescriptions in the strong-field, as will be discussed in the
course of this paper.

We analyze and contrast, for the first time, the dynamics and associated GW fluxes
obtained when using the
\begin{enumerate}
 \item[i)]~MP with the Tulczyjew (T) SSC \cite{tulczyjew1959motion}, 
 \item[ii)]~MP with the Pirani (P) SSC \cite{Pirani:1956tn},
 \item[iii)]~MP with the Ohashi-Kyrian-Semerak (OKS) SSC \cite{Ohashi:2003,Kyrian:2007zz},
 \item[iv)]~Hamiltonian EOM based on an effective-one-body Hamiltonian
   with the linearised T SSC \cite{Damour:2014sva}.
\end{enumerate}
We compare gauge-invariant energetics of circular orbits and
the ISCO frequencies. We find that all the dynamics are compatible
in terms of the energetics, shown in Fig's~\ref{fig:Ex}-\ref{fig:eSO},
for dimensionless particle spins with absolute value less than $0.2$,
where a spin with value $1$ corresponds to the extremal case when viewing the body
as a spinning BH. 
In this regime of small spins the relative differences in the ISCO shifts
are below $20\%$, see Fig \ref{fig:xISCO_shift}.
Additionally, we compute the respective asymptotic GW fluxes
at null-infinity using the time-domain Teukolsky-approach of
\cite{Harms:2015ixa}, hereafter Paper~\RM{1}.
As shown in Fig~\ref{fig:F_lsum_m123_over_x_differentEOMs},
we find that the GW fluxes relative to the different dynamics agree
with each other within our numerical precision at low frequencies (large orbital
radii). At high frequencies, i.e.\ small orbital radii close to the
respective ISCOs our results, however, indicate that the different
dynamics yield significantly different GW fluxes.

The article is organized as follows. In Sec~\ref{sec:MP_dynamics}
we review the MP formalism and the SSCs employed in this work.
For all the cases we work out how circular equatorial orbits
(CEOs) and the ISCO locations can be found numerically.
Similarly, in Sec~\ref{sec:HamDynamics} we
review the Hamiltonian formalism, and the corresponding CEO and ISCO computations.
In Sec~\ref{sec:dynamics_energeticsANDisco} we analyze the circular dynamics given by
the different EOM using both binding energy curves and the
spin-dependent shift of the ISCO frequency. 
In Sec~\ref{sec:comparison_fluxes} we compare the asymptotic GW
fluxes computed at null-infinity.

\paragraph*{Units and notation:}
Geometric units are used throughout the work, ${G=c=1}$.
We use the Riemann tensor defined as
${{R^\alpha}_{\beta\gamma\delta}=
 \Gamma^\alpha_{\gamma \lambda} \Gamma^\lambda_{\delta \beta}
 - \partial_\delta \Gamma^\alpha_{\gamma\beta}
 - \Gamma^\alpha_{\delta\lambda} \Gamma^{\lambda}_{\gamma\beta}
 + \partial_\gamma \Gamma^{\alpha}_{\delta \beta}}$,
where the Christoffel symbols $\Gamma$ are computed
from the metric with signature $(-,+,+,+)$. 

We describe a generic binary system
with the two masses $m_1,m_2$ and spin magnitudes $S_1,S_2$
in the convention that $m_1 \gtrsim m_2$.
We define $M\equiv m_1+m_2$, $\mu \equiv m_1 m_2/M$ and $\nu\equiv \mu/M$.
The test-particle limit, which we consider in the numerical experiments of this work,
is then understood by $M = m_1 \gg m_2=\mu$
and $\nu=0$. More precisely, in the test-particle limit we denote by 
$M$ the mass of the central BH and by $\mu$ the mass of the particle.

In fact, for a spinning particle there are different notions of mass.
In most cases considered here the conserved mass of the particle is defined as $\mu:= \sqrt{-p^\mu p_\mu}$.
Exceptionally, when discussing the MP with the P SSC in Sec~\ref{sec:MP_dynamics_CEOs},
the conserved rest-mass is differently defined, see Eq~\eqref{eq:fatm_mass}, and called $\textsf{m}$. 
Note that, when using dimensionless quantities that involve the particle's rest mass,
one would thus have to use different quantities for the different dynamics,
to be strict (Appendix~\ref{sec:solution_methods}). For simplicity we will not do so but use the same
symbols and expect that this subtlety is understood by the reader;
e.g.\ we always denote the dimensionless spin of the particle by $\sigma$,
which means $\sigma\equiv S_2/(\textsf{m} M)$ for the P SSC
but $\sigma\equiv S_2/(\mu M)$ for the other cases.  
In practice, in the perturbative calculations both $\mu$ and $M$
scale away so that we work numerically with $\mu=M=1$,
and the variable $\sigma$ varies between $-1\leq \sigma \leq 1$.

For the Kerr BH background with spin angular momentum $S_1=a_1 M = \hat{a}_1 M^2$,
and its nonspinning Schwarzschild limit,
we use the standard Boyer-Lindquist (BL) coordinates
$\{t,r,\theta,\phi\}$.
In the transition of the EOB description for generic binaries
to the extreme-mass-ratio limit in Sec~\ref{sec:HamDynamics}
we also need the EOB radial coordinate of the
deformed Kerr background, $r^{\rm{EOB}}$, and we denote its mass-reduced form by $ \hat{R}=r^{\rm{EOB}}/M$.
Note that $r^{\rm{EOB}}=r$ only if $\nu=0$ and $S_1=0$, $S_2=0$, i.e.\ 
even for $\nu=0,S_1=0$ the background is deformed if $S_2 \neq 0$,
cf.~Sec~\ref{sec:dynamics_energeticsANDisco} for more details.

\section{Mathisson-Papapetrou dynamics}
\label{sec:MP_dynamics}

In the following section we review the MP and
discuss how circular orbits can be produced
numerically using either the MP with the T SSC, the P SSC, or the OKS SSC. 
The analogue discussion for the Hamiltonian approach is given in Sec~\ref{sec:HamDynamics}.

\subsection{EOM and SSC}
\label{sbsec:MP_EOM_SSC}

The MP in their nowadays standard form read~\cite{dixon1974dynamics}
\begin{subequations}
\label{eq:MP}
\begin{align}
\frac{D~p^{\mu}}{d\lambda}&=-\frac{1}{2}~{R^{\mu}}_{\nu\kappa\lambda}v^{\nu}S^{\kappa\lambda}
\ , \label{eq:MPp}\\
\frac{D~S^{\mu\nu}}{d\lambda}&=p^{\mu}~v^{\nu}-v^{\mu}~p^{\nu} \ ,\label{eq:MPS}
\end{align}
\end{subequations}
where $D/d\lambda \equiv v^\mu \nabla_\mu$.
The system of equations~\eqref{eq:MP} is not closed, and a SSC must be
specified in order to confine to a unique solution.

A common procedure for finding a SSC stems from our
Newtonian intuition that \textit{spin} should be space-like,
that is, the spin-tensor should be orthogonal to the four-velocity 
of some preferred time-like observer.
Representing this observer by some future-pointing time-like
vector $V^\mu$ with 
\be \label{eq:normV}
V^\mu V_\mu=-1 \ ,
\ee
the general form of a SSC reads
\be \label{eq:SSC}
V_\mu S^{\mu\nu}=0 \ .
\ee
Three of the four conditions~\eqref{eq:SSC} are linearly
independent, and along with
Eqs~\eqref{eq:normV} they fix the centroid that is tracked by the MP.
For example, for the T SSC one takes $V^\mu=p^\mu / \mu$ (see below), where $\mu$
is the dynamical rest mass 
\begin{align}
 \label{eq:mu_mass}
 \mu:=\sqrt{-p^\mu p_\mu} \ .
\end{align}
For later reference, we also introduce here another notion of mass,
\begin{align}
 \label{eq:fatm_mass}
 \textsf{m}:=-v^\mu p_\mu \; ,
\end{align}
which is important for the P SSC.

In general, the particle's four-momentum $p_\mu$ 
and four-velocity $v^\mu$ are not parallel.
If they were, we would have $\frac{D~S^{\mu\nu}}{d\lambda}=0$
from Eq~\eqref{eq:MPS}.
In fact, rearranging that equation, one gets
\be \label{eq:hidmom}
 p^\mu=\textsf{m} v^\mu-v_\nu \frac{D~S^{\mu\nu}}{d\lambda} \qquad ,
\ee
where the second term is known as the \textit{hidden} momentum, i.e.
\be
p^\mu_{\rm{hidden}} := p^\mu-\textsf{m} v^\mu  
\ee
(see, e.g., \cite{Semerak:2015dza,Gralla:2010xg}). 
As discussed below, the OKS SSC is characterized by $p^\mu_{\rm{hidden}}=0$.

Having defined the observer's reference vector $V^\mu$,
it is possible to introduce the spin four-vector 
\begin{align} \label{eq:SpinVect}
 S_\mu = -\frac{1}{2} \epsilon_{\mu\nu\rho\sigma}
          \, V^\nu \, S^{\rho\sigma} \qquad,
\end{align}
where $\epsilon_{\mu\nu\rho\sigma}=\sqrt{-g} \tilde{\epsilon}_{\mu\nu\rho\sigma} $
is the Levi-Civita tensor with the Levi-Civita symbol
$\tilde{\epsilon}_{0123}=1$ and the determinant $g$ of the background metric
tensor. The inversion of Eq~\eqref{eq:SpinVect} reads
\begin{align}    \label{eq:T4VSin}
   S^{\rho\sigma}=-\epsilon^{\rho\sigma\gamma\delta} S_{\gamma}
   V_\delta \qquad ,
\end{align}
and the spin's magnitude is 
 \be \label{eq:spinMagnitude}
  S^2=\frac{1}{2}S^{\mu\nu}S_{\mu\nu}=S^\mu S_\mu \quad .
 \ee
The constancy of the scalar quantities $\mu,\textsf{m}$ and $S$
depends on the choice of the SSC, and it is summarized in
Tab~\ref{tab:SSC_constants}. For instance, for the T SSC, $\mu$ is
constant but $\textsf{m}$ is not, and vice versa for the P
SSC~\cite{Semerak:1999qc}. 
For the OKS SSC both notions of mass are constant~\cite{Kyrian:2007zz}.
In general, the spin-magnitude $S$ is not constant upon evolution,
but for all SSCs discussed here it is so
(see, e.g.,~\cite{Kyrian:2007zz,Semerak:1999qc}). 

Besides the SSC-dependent constants of motion, there
are more general, background-dependent constants constructed
from Killing vectors.
In particular, for a Killing vector $\xi^\mu$ the quantity 
\begin{align} \label{eq:ConsMot}
 C=\xi^\mu p_\mu-\frac12 \xi_{\mu;\nu} S^{\mu\nu}
\end{align}
remains conserved upon evolution \cite{Dixon:1970zza}.
For stationary and axisymmetric spacetimes with a reflection symmetry along the
equatorial plane (SAR spacetimes) we have, using BL-coordinates, the two Killing 
vector fields
$\xi^\mu_{(t)}={\delta^\mu}_t$ and $\xi^\mu_{(\phi)}={\delta^\mu}_\phi$. 
The corresponding conserved quantities are 
 \begin{align}\label{eq:EnCons}
  E &:= -p_t+\frac12g_{t\mu,\nu}S^{\mu\nu} \qquad,
 \end{align}
and
  \begin{align}\label{eq:JzCons}
  J_z &:= p_\phi-\frac12g_{\phi\mu,\nu}S^{\mu\nu} \qquad,
 \end{align}
respectively.
${E=\rm{const.}}$ corresponds to the conservation of energy and
${J_z=\rm{const.}}$ to the 
conservation of the component of the total angular momentum along the
symmetry axis $z$. \\

\begin{table}[t]
\caption{ 
Constancy of the two notions of mass $\mu$, Eq~\eqref{eq:mu_mass},
and $\textsf{m}$, Eq~\eqref{eq:fatm_mass}, and of the spin-magnitude $S$, Eq~\eqref{eq:spinMagnitude},
for the Tulczyjew (T) SSC, the Pirani (P) SSC, and the Ohashi-Kyrian-Semerak (OKS) SSC.
Constancy is denoted by a \checkmark, whereas $\times$
marks that the quantity is not constant.
For more details some references to appearances of these SSCs in the literature are included.
 }
\centering
  \begin{tabular}[t]{c || c | c c c | c } 
  { SSC } &
  $V^\mu$ &
  $\mu$ &
  $\textsf{m} $ &  
  $S$  &
  References \\
  \hline
  \hline
   T & Eq~\eqref{eq:V_TSSC} & \checkmark & $\times$ & \checkmark & \cite{tulczyjew1959motion, Semerak:1999qc,Kyrian:2007zz, Han:2014ana,Hackmann:2014tga,Lukes-Gerakopoulos:2014dma}  \\
   P & Eq~\eqref{eq:V_PSSC} & $\times$ & \checkmark & \checkmark  & \cite{Pirani:1956tn, Semerak:1999qc, Kyrian:2007zz, Hackmann:2014tga} \\
   OKS & Eq~\eqref{eq:V_OKSSSC} & \checkmark & \checkmark & \checkmark & \cite{Ohashi:2003,Kyrian:2007zz,Semerak:2015dza}  
  \end{tabular} 
\label{tab:SSC_constants}
\end{table}

In the following we briefly introduce the SSCs used in this work. A thorough
analysis of these conditions can be found in~\cite{Kyrian:2007zz} and \cite{Semerak:2015dza}.

\subsubsection{Tulczyjew SSC}
For the T SSC~\cite{tulczyjew1959motion} the reference vector is
\be
\label{eq:V_TSSC}
V^\mu=\frac{p^\mu}{\mu} \qquad .
\ee
This choice makes the spin spatial for an observer moving in the
direction of the four momentum,
\begin{align}
 \label{eq:T_SSC}
  p_\mu S^{\mu\nu}= 0  \qquad \text{(T SSC)} \ .
\end{align}
For the T SSC an explicit relation between $v^\mu$ and $p_\mu$,
$S^{\mu\nu}$ can be found, i.e.
\begin{align} \label{eq:v_p_TUL}
 v^\mu = \frac{\textsf{m}}{\mu^2} \left(
          p^\mu + 
          \frac{ 2 \; S^{\mu\nu} R_{\nu\rho\kappa\lambda} p^\rho S^{\kappa\lambda}}
          {4 \mu^2 + R_{\alpha\beta\gamma\delta} S^{\alpha\beta} S^{\gamma\delta} }
          \right)  \quad .
\end{align}
The T SSC is widely used in numerical applications, e.g.~\cite{Saijo:1998mn,Hartl:2003da,Han:2010tp,Hackmann:2014tga}.
In particular, in Paper~\RM{1}, we have already discussed the T SSC
(cf. Sec~\RM{2}~C therein) and computed the GW fluxes produced by a spinning particle in circular orbits.\\

\subsubsection{Pirani SSC}
For the P~SSC~\cite{Pirani:1956tn} the reference vector is the four-velocity, i.e. 
\be 
\label{eq:V_PSSC}
V^\mu=v^\mu \qquad ,
\ee
making spin spatial for an observer moving in the direction of the particle's four-velocity,
\begin{align}
 \label{eq:P_SSC}
  v_\mu S^{\mu\nu}= 0 \qquad \text{(P SSC)}  \ .
\end{align}
Note that sometimes this choice is called the ``Frenkel'' SSC~\cite{Hackmann:2014tga}.
The evolution equation of the four-velocity for the P SSC is given by
\be \label{eq:veloc_evol_P}
 \frac{D v^\mu}{d \lambda}=-\frac{1}{S^2}\left(\frac{A}{2 \textsf{m}}S^\mu+p_\kappa S^{\mu \kappa}\right) \qquad,
\ee
where
\be
\label{eq:A}
 A=R_{\mu\nu\kappa\lambda}S^\mu v^\nu S^{\kappa \lambda} \qquad.
\ee

For the derivation of the equation see, e.g., \cite{Kyrian:2007zz}.\\

\subsubsection{Ohashi-Kyrian-Semerak SSC} \label{sec:OKS}
The OKS SSC was proposed in~\cite{Ohashi:2003,Kyrian:2007zz} 
and revised recently in~\cite{Semerak:2015dza}; we work with the latter
version.  The main idea of the OKS SSC is to exploit the freedom in the choice of
the future-pointing time-like vector $V^\mu$ to impose desirable features in the 
EOM,  like the cancellation of the hidden momentum.
To fulfill the OKS SSC upon evolution, one promotes $V^\mu$ to an evolution variable of
the system and deduces an evolution equation for it. The latter is then solved 
with suitable initial data. 

In the OKS framework the covariant
derivative of the timelike four vector, $D V^\mu/ d \lambda$, has to be
proportional to $S^\mu$ and satisfy the condition
\be
\label{eq:V_OKSSSC}
\frac{D V_\mu}{d \lambda} S^{\mu\nu}=0 \qquad \text{(OKS SSC)} \ .
\ee
Of course, $V_\mu S^{\mu\nu}=0$ has to hold as well.
The condition \eqref{eq:V_OKSSSC}
eliminates the hidden momentum, and therefore 
\be
 p^\mu=\textsf{m} v^\mu \ .
\ee
Once the latter holds, it is straightforward that 
\be
\mu=\textsf{m} . 
\ee

According to \cite{Semerak:2015dza} a ``natural'' way to restrict the possible
$V^\mu$ is to require $\frac{D( v^\mu S_\mu) }{d \lambda}=0$. 
Then one can deduce the following evolution equations for $V^\mu$
\be \label{eq:OKS_Veom}
 \frac{D V^\mu}{d \lambda}=\frac{\alpha}{\textsf{m} \mu^2} S^\mu \qquad,
\ee
and for $S^\mu$
\be \label{eq:OKS_Seom}
 \frac{D S^\mu}{d \lambda}=\frac{\alpha S^2}{\textsf{m} \mu^2} V^\mu \qquad,
\ee
respectively, where
\be
\label{eq:alpha}
\alpha=\frac{\mu^2}{S^2}\frac{D p^\mu}{d \lambda} S_\mu \qquad.
\ee
For the derivation of Eqs~\eqref{eq:OKS_Veom},~\eqref{eq:OKS_Seom}, see~\cite{Semerak:2015dza}.
Note that the OKS SSC does not specify a unique worldline itself
unless an initial $V^\mu$ has been set.

\subsection{Circular equatorial orbits (CEOs)}
\label{sec:MP_dynamics_CEOs}

The problem of finding CEOs reduces to select 
appropriate initial data for the variables~$\{X^\mu, v^\mu,p_\mu, S^{\mu\nu}\}$
so that circular equatorial motion is obtained upon evolution of the MP.
In the following, we describe methods for producing such CEO initial data valid for
arbitrary SAR spacetimes. We first
discuss the part that is common to all the SSCs and then specify the details for
each choice of SSC. Additionally, for each SSC we discuss how to
determine the ISCO. The latter problem is nontrivial because, as we
shall see, the relations between $v^\mu$ and $p_\mu$ and the constants
of motion $E$ and $J_z$ can become complicated. A new procedure to
find CEOs and ISCOs is presented for the P and the OKS SSC.
Before going into detail, note that, in order to allow eternal circular motion,
our setup completely neglects self-force effects
of the small body, which like the spin-curvature coupling of the MP
in principle lead to deviations from geodesic motion~\cite{Ruangsri:2015cvg}.

\paragraph*{Coordinates:}
Without loss of generality, we identify the time coordinate of the
particle with the background coordinate time $t$. 
For the spatial coordinates we set the initial data according to the
assumptions that 
\begin{align}
 r        =\rm{const.} \; , \quad 
 \theta   = \frac{ \pi } {2} \; , \quad
 \phi     =\Omega \, t  \;  .
 \label{eq:assumptions_coordinates}
\end{align}
Here $\Omega \equiv d\phi/dt$ is the orbital frequency of the particle,
which is expected to remain constant during the evolution, and 
whose initial data will be determined below from the tangent vector.
Note that in Eq~\eqref{eq:assumptions_coordinates}
we are not introducing a distinct notation for the particle's coordinates
and the background coordinates respectively; e.g., we simply write $r$
for the particle's BL radius, assuming that the meaning is always
comprehensible from the context.

\paragraph*{Tangent vector:}
It is clear that for CEOs we need
\begin{align}
  v^r=0, \quad v^\theta=0 \; ,
  \label{eq:assumptions_tangent}
\end{align}
which we therefore set in the initial data.
It is not trivial though how $v^t(t=0)$ and $v^\phi(t=0)$ should be determined.
In fact, it turns out that the procedures depend on the choice of the SSC
and will therefore be discussed separately for each SSC below.
Nevertheless, since the time coordinate of the particle is identified 
with the background coordinate time, $v^t$ is just the lapse of the
particle and the relation
\begin{align}
 \label{eq:lapse}
 v^t = \frac{1}{\sqrt{-g_{tt} - 2 g_{t\phi} \Omega - g_{\phi\phi} \Omega^2}}
\end{align}
is always fulfilled for our calculations.

\paragraph*{Four momentum:}
The treatment of $p^\mu$ depends on the choice of the SSC
and it is not always necessary to make additional assumptions on the momenta.
For example, if a SSC entails $p^\mu \parallel v^\mu$,
Eq~\eqref{eq:assumptions_tangent}  
already implies that the momenta in the radial and polar directions vanish.
As a matter of fact, inspecting our dynamical data \textit{a posteriori},
we find that all SSCs tested here share the common feature that 
\begin{align}
 p_r = 0, \quad p_\theta = 0 \ .
  \label{eq:ID_4mom}
\end{align}
The reasons are discussed below for each SSC separately.

\paragraph*{Spin tensor:}
We demand that the spin-vector of the particle is aligned
with the orbital angular momentum, 
\begin{align}\label{eq:Spintheta}
  S^\mu = S^\theta {\delta^\mu}_\theta \ .
\end{align}
Note that Eq~\eqref{eq:Spintheta} combined with
Eqs~\eqref{eq:assumptions_tangent} imply that the condition, 
\begin{align} 
 \label{eq:v_perp_Svec}
 v^\mu S_\mu=0 \; ,
\end{align}
is met for all three SSCs that we consider. 
This is obvious for the P SSC, but it can be shown also for the T SSC,
e.g.~\cite{Hackmann:2014tga}, and it can be demanded for the OKS
SSC.  

The spin-vector~\eqref{eq:Spintheta} can be expressed through the
spin-magnitude $S$ using Eq~\eqref{eq:spinMagnitude}, which gives
\be 
S_\theta=-\sqrt{g_{\theta\theta}}~S \qquad,
\ee
with $S>0$ ($S<0$) corresponding to a spin-vector that is (anti-)aligned with the 
orbital angular momentum, which by convention is always pointing along the positive z-direction
in our setup.  

Inserting assumption~\eqref{eq:Spintheta} into Eq~\eqref{eq:T4VSin}, we get a
general prescription for setting the spin-tensor
\begin{subequations}
\label{eq:SpinTensEQ}
 \begin{align} 
  S^{tr} &= -S~V_\phi \sqrt{-\frac{g_{\theta\theta}}{g}}= -S^{rt}  \qquad, \\
  S^{t\phi} &= S~V_r \sqrt{-\frac{g_{\theta\theta}}{g}}= -S^{\phi t}
  \qquad,  \\
  S^{r\phi} &= -S~V_t \sqrt{-\frac{g_{\theta\theta}}{g}}= -S^{\phi r}  \; ,
 \end{align}
\end{subequations}
which we use for all three SSCs by replacing the vector $V_\mu$ accordingly, see
Sec~\ref{sbsec:MP_EOM_SSC}. Note that at this stage the initial data of
$V_\mu$ for the OKS SSC are still missing, and they are discussed in
Sec~\ref{sec:CEOs_SSC_OKS}.

\paragraph*{Energy and angular momentum constants:}
With the relations~\eqref{eq:SpinTensEQ}
the constants $E$ and $J_z$, given by Eqs~\eqref{eq:EnCons} and~\eqref{eq:JzCons},
can be written as
\begin{align}\label{eq:ConEq_J}
J_z &=p_\phi+\frac{\sqrt{g_{\theta\theta}}}{2\sqrt{-g}} S(g_{t \phi,r}
V_\phi-g_{\phi\phi,r} V_t) \ , \\
\label{eq:ConEq_E}
E &=-p_t+\frac{\sqrt{g_{\theta\theta}}}{2\sqrt{-g}}S( g_{t \phi,r}
V_t- g_{t t,r}V_\phi) \ .
\end{align} 
Using these equations we are able to specify initial data for $(E,J_z)$
instead of $(p_t,p_\phi)$. \\

The procedures to set the remaining initial conditions for CEOs is now discussed
separately for each SSC.

\subsubsection{Tulczyjew SSC}
\label{sec:CEOs_SSC_T}

To find CEOs under the T SSC, one replaces ${V^\mu=p^\mu/\mu}$ in Eqs~\eqref{eq:ConEq_J},
and rearranges the equations such that $p_t$ and $p_\phi$ are functions of
$E,~J_z,~r$. The rearranged equations are plugged into Eq.~\eqref{eq:mu_mass} to obtain $p_r$ as function of $E,~J_z,~r$. Inserting
the above rearranged components of $p_\mu$ in Eq~\eqref{eq:v_p_TUL} for
the radial component of the tangent vector, one gets
\begin{align}
 \label{eq:VeffT}
 v^r \propto \sqrt{V_{\rm{eff,T}}} \; ,
\end{align}
where the function $V_{\rm{eff,T}}=V_{\rm{eff,T}}(E,~J_z,~S,~r)$ is 
an \textit{effective potential}, by analogy with the effective
potential used for a nonspinning particle, see
Appendix~\ref{sec:ISCO_nonspinning}.
The explicit form of $V_{\rm{eff,T}}$ can be found in Eq~(20) of Paper I.

Motion can take place only when $V_\textrm{eff,T}\geq 0$. 
For $V_\textrm{eff,T}=0$ one gets the turning points of the motion in radial
direction. However, CEOs have fixed radii, which means that the turning points
should be also extrema of $V_\textrm{eff,T}$. Thus, for a CEO it holds that 
\begin{align} \label{eq:VeffCEO_T}
  V_\textrm{eff,T} = 0, \qquad \frac{d V_\textrm{eff,T}}{d r} = 0   \qquad .
\end{align}
The solution of the system~\eqref{eq:VeffCEO_T} for a given radial distance $r$
and spin $S$ provides the energy $E$ and the z-component of the total
angular momentum $J_z$. For the Kerr background
the solution of the system~\eqref{eq:VeffCEO_T} has been found analytically,
see, e.g., \cite{Hackmann:2014tga,Tanaka:1996ht}. 

An ISCO is a CEO located at an inflection point of the effective potential, in
our case of $V_\textrm{eff,T}$.
Thus,
to find an ISCO's $r$, $E$, and $J_z$ for a given spin $S$, we solve
the system~\eqref{eq:VeffCEO_T} along with the condition 
$\frac{d^2 V_\textrm{eff,T}}{d r^2} = 0$.

\subsubsection{Pirani SSC}
\label{sec:CEOs_SSC_P}

In contrast to the single effective potential used to find CEOs for the T
SSC, the CEOs for the P SSC are determined here using three ``potentials''
named $V_\textrm{P}$, $V_{\textrm{eff},P}$, and $V_{\textrm{con},P}$. 

For equatorial motion in a SAR spacetime it holds that $A=0$
(cf.~Eq~\eqref{eq:A}). Furthermore, once 
we demand $v^r=0$ and $p^r=0$, Eq~\eqref{eq:veloc_evol_P} implies that
the polar acceleration vanishes, $d v^\theta/d \lambda=0$, as well as the time
component $d v^t/d \lambda$ and the azimuthal component $d v^\phi/d \lambda$.
The radial component of Eq~\eqref{eq:veloc_evol_P} is reduced to
\be
 \frac{d v^r}{d \lambda}=-\frac{V_\textrm{P}}{2 S~g_{rr}~\sqrt{-g} } \qquad,
\ee
where we define 
 \begin{widetext}
\be
 V_\textrm{P}:=2 g_{rr} \sqrt{g_{\theta\theta}}\left(g_{\phi\phi}v^{\phi}p_t-g_{tt}v^t p_\phi+g_{t\phi}(v^t p_t-v^\phi p_\phi)\right)-S\sqrt{-g}
 \left(\frac{\partial g_{tt}}{\partial r}{v^t}^2+2 v^t v^\phi \frac{\partial g_{t\phi}}{\partial r}+\frac{\partial g_{\phi\phi}}{\partial r}{v^\phi}^2\right) \qquad,
\ee
 \end{widetext}
in which $p_t$ and $p_\phi$ are replaced using
Eqs~\eqref{eq:ConEq_J}. The condition  $V_\textrm{P}=0$ prevents
radial acceleration.  

As for the T SSC, the four-velocity contraction provides an effective potential,
$V_{\textrm{eff},P}$. Rearranging ${v^\mu v_\mu=-1}$ so to
express the radial velocity $v^r$ as a function of $v^t,~v^\phi$ and $r$, one gets 
\be
 v^r=\pm\sqrt{\frac{V_{\textrm{eff},P}}{g_{rr}}}   \qquad ,
\ee
where
\be
 V_{\textrm{eff},P}:=-(g_{tt} {v^t}^2+2 g_{t \phi } v^t
 v^\phi+g_{\phi\phi} {v^\phi}^2+1)  \qquad . 
\ee 
Motion is allowed only when $V_{\textrm{eff},P} \geq 0$. 

Finally, by rewriting the definition of the mass $\textsf{m}$ to express the
function $v^r p_r$ in terms of $v^t,~v^\phi,~p_t,~p_\phi$, one
obtains 
\be
 V_\textrm{con,P}:=v^r p_r=-\textsf{m}-v^t p_t-v^\phi p_\phi \qquad .
\ee
In the above expression $p_t$ and $p_\phi$ are replaced by Eqs~\eqref{eq:ConEq_J},
in which for the P SSC $V^t=v^t$, $V^\phi=v^\phi$. Thus,
$V_\textrm{con,P}=V_\textrm{con,P}(v^t,~v^\phi,~r,~S,~E,~J_z)$, which
gives the third potential.

To find CEOs for a given $r$ and $S$, we solve the
system
\begin{align} \label{eq:PCEOs} 
 V_\textrm{eff,P}=0, \quad V_\textrm{P}=0, \quad V_\textrm{con,P}=0 \qquad,\nonumber \\
 \frac{d V_\textrm{eff,P}}{d r}=0, \quad 
 \frac{d V_\textrm{P}}{d r}=0, \quad 
 \frac{d V_\textrm{con,P}}{d r}=0 \qquad,
\end{align}
where we consider $v^t,~v^\phi$ as functions of $r$.
Thus, in the latter system the variables are 
$v^t,~v^\phi,~d v^t/d r,~d v^\phi/dr,~\hat{E},~\hat{J}_z$.
As far as we know, there are no analytical solutions for CEOs in the literature
for the P SSC, and the above numerical procedure is novel. Note that our numerical findings
show that this procedure can avoid the helical motion appearing in studies that
use the MP with the P SSC~\cite{Kyrian:2007zz}. 
Helical motion was the reason that the P SSC was considered unphysical
for a long time. This misconception has been explained in 
\cite{Costa:2012}, where it has been shown that the P SSC is physically acceptable. 

Here, the ISCO can be found by searching for inflection points of the three potentials.
For a given $S$ we solve the system~\eqref{eq:PCEOs} plus 
\be \label{eq:PISCOs} 
\frac{d^2 V_\textrm{eff,P}}{d r^2}=0, \quad 
\frac{d^2 V_\textrm{P}}{d r^2}=0, \quad 
\frac{d^2 V_\textrm{con,P}}{d r^2}=0 \qquad,
\ee
where the variables are $r,~v^t,~v^\phi,~d v^t/d r~d v^t/d r,~d v^\phi/d r$, 
$d v^\phi/d r,~d^2 v^t/d r^2,~d^2 v^\phi/d r^2,~E,~J_z$.

\subsubsection{Ohashi-Kyrian-Semerak SSC}
\label{sec:CEOs_SSC_OKS}

For the OKS SSC there is no procedure in the literature describing how
to find CEOs. We present here a working solution,
following the ideas applied for the T SSC and the P SSC. We note that our general
demands for CEOs, Eqs~\eqref{eq:assumptions_tangent} and~\eqref{eq:Spintheta}, 
are compatible with the OKS condition \eqref{eq:V_OKSSSC}.
To see this, note that Eqs~\eqref{eq:assumptions_tangent}
and~\eqref{eq:Spintheta} already imply $v^\mu S_\mu=0$, 
which is even stronger than the OKS requirement $\frac{D( v^\mu S_\mu) }{d \lambda}=0$. 

For equatorial motion in a SAR spacetime it holds that ${\alpha=0}$
(cf.~Eq~\eqref{eq:alpha}). Since ${\alpha=0}$,
Eqs~\eqref{eq:OKS_Veom},~\eqref{eq:OKS_Seom} imply that $V^\mu$ and
$S^\mu$ are parallel transported along the worldline, i.e.
\be \label{eq:OKS_Veom_eq}
 \frac{D V^\mu}{d \lambda}=0 \qquad,
\ee 
and
\be \label{eq:OKS_Seom_eq}
 \frac{D S^\mu}{d \lambda}=0 \qquad.
\ee 

Before we proceed, the conservation of the ansatz~\eqref{eq:Spintheta} in time has to be
checked for the OKS SSC. 
From Eq~\eqref{eq:OKS_Seom_eq} we get identities $0=0$, apart from the
$\theta$ component which reads  
\begin{align}
 \frac{d S^\theta}{d \lambda}+\frac{S^\theta v^r}{2 g_{\theta\theta}}
 \frac{\partial g_{\theta\theta}}{\partial r}=0 \qquad.
\end{align}
From the latter we confirm that if $v^r=0$, then $S^\theta$ is constant. Thus, our ansatz holds upon evolution also for the OKS SSC.

As discussed in Sec~\ref{sec:OKS}, we can exploit the fact that $V^\mu$ is
a relatively arbitrary future-pointing time-like vector in order to 
get desired features, and this is what we do in order to get CEOs.
First we note that the $\theta$ component of Eq~\eqref{eq:OKS_Veom_eq} gives 
\begin{align}
\frac{d V^\theta}{d \lambda}+\frac{V^\theta v^r}{2 g_{\theta\theta}}
 \frac{\partial g_{\theta\theta}}{\partial r}=0 \qquad. \nonumber
\end{align}
We can simply impose 
\begin{align}
 \label{eq:ansatz_Vtheta_OKS}
 V^\theta=0 \; ,
\end{align}
which is certainly a natural choice for CEOs since the motion takes place on the
equatorial plane.

As in the case of the P SSC, we have to use three ``potentials''. The effective potential
comes from the four-velocity contraction, or for the OKS SSC equivalently from the
four-momentum contraction $p^\mu p_\mu=-\mu^2$. Namely,
\be
 p_r=\pm\sqrt{\frac{V_\textrm{eff,OKS}}{g^{rr}}} \; ,
\ee
where
\be \label{eq:Veff_OKS}
 V_\textrm{eff,OKS}=-(\mu^2+{p_t}^2 g^{tt}+{p_\phi}^2 g^{\phi\phi}+2 p_t p_\phi g^{t\phi}) \qquad.
\ee 
In Eq~\eqref{eq:Veff_OKS} $p_t$ and $p_\phi$ have to be replaced using Eqs~\eqref{eq:ConEq_J}
in order to make $V_\textrm{eff,OKS}$ a function of
$r,~S,V_t,~V_\phi,~E,~J_z$. Notably, $V_\textrm{eff,OKS}$
does not depend on the $V_r$ component, which means that for simplicity
we can set 
\begin{align}
 \label{eq:ansatz_Vr_OKS}
 V_r=0 \; .
\end{align}
In fact this requirement is rather convenient since it gives
us a relation between $V_t$ and $V_\phi$ through the fact that $V^\mu V_\nu=-1$
(recall that we have set $V_\theta=0$), i.e.\
\be \label{eq:V_OKS1}
 V_\textrm{OKS1}=1+g^{tt}{V_t}^2+g^{\phi\phi}{V_\phi}^2+2 g^{t\phi} V_t V_\phi=0 \qquad ,
\ee
and, thus, we have the second potential.
The requirements $V_r=0$ and $v^r=0$ reduce the time and the
azimuthal components of Eq~\eqref{eq:OKS_Veom_eq} to ${\frac{D V_t}{d \lambda}=0}$
and ${\frac{D V_\phi}{d \lambda}=0}$, while the radial component gives us the
third potential. Namely,
\be 
 \frac{d V_r}{d \lambda}=\frac{V_\textrm{OKS2}}{2 \mu ({g_{t\phi}}^2- g_{\phi\phi} g_{tt})^2} \qquad,
\ee
where
\begin{widetext}
\begin{align} \label{eq:V_OKS2}
 V_\textrm{OKS2} &=V_\phi\left((g_{t\phi} p_t-g_{tt}p_\phi)(g_{t\phi}\frac{\partial g_{t\phi}}{\partial r}-g_{tt}\frac{\partial g_{\phi\phi}}{\partial r})
-(g_{\phi\phi} p_t-g_{t\phi}p_\phi)(g_{t\phi}\frac{\partial g_{tt}}{\partial r}-g_{tt}\frac{\partial g_{t\phi}}{\partial r})\right)  \nonumber \\
  &+V_t \left((g_{\phi\phi} p_t-g_{t\phi}p_\phi)(g_{\phi\phi}\frac{\partial g_{tt}}{\partial r}-g_{t\phi}\frac{\partial g_{t\phi}}{\partial r})
-(g_{t\phi} p_t-g_{tt}p_\phi)(g_{\phi\phi}\frac{\partial g_{t\phi}}{\partial r}-g_{t\phi}\frac{\partial g_{\phi\phi}}{\partial r})\right)   \qquad,
\end{align}
\end{widetext}
which we demand to be zero, i.e. we demand that $d V_r/d \lambda=0$ 
so that $V_r=0$ remains satisfied in time.
    
To find the turning points for a given radius $r$ 
and spin $S$, one has to solve the system
\be \label{eq:OKS_tp}
 V_\textrm{eff,OKS}=0, \quad V_\textrm{OKS1}=0, \quad V_\textrm{OKS2}=0 \quad,
\ee 
so that one of the variables $V_t,~V_\phi,~E,~J_z$ can be expressed
as a function of the other three.
The procedures to find CEOs and the ISCOs are similar to the procedures
described in Sec~\ref{sec:CEOs_SSC_P} for the P SSC. In particular, we find
CEOs for a given $r$ and $S$ by solving the system of the three potentials and
their first derivatives with respect to $r$, where the unknowns are
$V_t,~V_\phi,~d V_t/dr,~d V_\phi/dr,~E,~J_z$. The ISCOs are found for a given $S$
by solving the system of the three potentials and their first and second
derivatives with respect to $r$, where the unknowns are
$V_t,~V_\phi,~d V_t/dr,~d V_\phi/dr,~d^2 V_t/dr^2,~d^2 V_\phi/dr^2,~E,~J_z$.

\section{Circular Dynamics of the Effective-One-Body Hamiltonian }
\label{sec:HamDynamics}

The Hamiltonian of a spinning particle on a Kerr BH,
written in a certain coordinate system and spin gauge, 
was obtained in Ref.~\cite{Barausse:2009aa} at linear order in the spin,
and it has been used, for example, to incorporate spin effects to the EOB model of~\cite{Pan:2013rra,Taracchini:2013rva, Babak:2016tgq}.
Recently that Hamiltonian was improved to quadratic order in Ref.~\cite{Vines:2016unv}.
By contrast, here we shall use the spinning-particle Hamiltonian obtained 
from the effective-one-body (EOB) Hamiltonian for nonprecessing spinning BHs 
of masses $m_1,m_2$ and dimensionful spins $S_1,S_2$ as introduced 
in Ref.~\cite{Damour:2014sva} (see Ref.~\cite{Balmelli:2015zsa} for the precessing version).
For application to our test-particle setup we
consider the following limits: (i) one body is much heavier than the 
other one, $m_1\gg m_2$; (ii) the heavier body is 
nonspinning, $S_1=0$; (iii) we restrict to circular dynamics;
(iv) we only consider spin-orbit couplings and drop
spin-spin ones.  
This yields a rather simplified description of the dynamics, 
since the EOB Hamiltonian of Ref.~\cite{Damour:2014sva} 
includes both spin-orbit (odd-in-spin) and spin-spin (even-in-spin) 
interactions in a resummed form, i.e. it incorporates an infinite 
number of spin-spin and spin-orbit couplings. We anticipate in 
passing that in another recently ongoing work, which will be 
published elsewhere, the equivalence between the circular dynamics 
entailed by the Hamiltonians of Ref.~\cite{Barausse:2009aa,Vines:2016unv} 
and the EOB-based Hamiltonian used here was checked explicitly.

For completeness, let us recall the structure of the complete EOB Hamiltonian
and how to get the spinning test-particle limit from it (see also Sec~II of Ref.~\cite{Bini:2015xua}).
We warn the reader that in the following several EOB-related quantities
will be introduced without detailed explanation nor discussion. 
This is done because the spinning test-particle limit of the 
EOB Hamiltonian of Ref.~\cite{Damour:2014sva} was not previously 
studied in our context and we think it is pedagogically useful to derive 
it from first principles, thus allowing an easy generalization to the Kerr case in the future.
Using the notation $M=m_1+m_2$, 
$\mu=m_1 m_2/M$, $\nu=\mu/M$, it is first written as
\be
H = M\sqrt{1+2\nu\left(\dfrac{H_{\rm eff}}{\mu}-1\right)} \; ,
\ee
where the effective Hamiltonian $H_{\rm eff}$ for equatorial dynamics
of parallel-spin binary systems reads
\be
H_{\rm eff}\equiv H_{\rm eff}^{\rm SO}+H_{\rm eff}^{\rm orb}=G^{\rm phys}_S P_{\phi} S + G_{S_*}^{\rm phys}P_{\phi} S_{*} + H_{\rm orb}^{\rm eff} \; ,
\ee
where $P_\phi$ is the total orbital angular momentum of the system,
while $S$ and $S_*$ are the following symmetric combinations of the two spins
\begin{align}
\label{eq:defS}
S&\equiv S_1 + S_2=m_1 a_1 + m_2 a_2\nonumber\\
&=m_1^2\hat{a}_1 + m_2^2\hat{a}_2 \; , \\
\label{eq:defSs}
S_*&\equiv \dfrac{m_2}{m_1}S_1+\dfrac{m_1}{m_2}S_2=m_2 a_1 + m_1 a_2\nonumber\\
 &=m_1m_2(\hat{a}_2+\hat{a}_2) \; .
\end{align}
The orbital effective Hamiltonian is
\be
\label{eq:Horb}
H_{\rm orb}^{\rm eff}=\sqrt{A\left(\mu^2+\dfrac{L^2}{r_c^2}+Q_4\right)+P_{r^{\rm EOB}_*}^2} \; ,
\ee
where $r_c^2$ is the squared centrifugal radius (that encodes here for
simplicity only leading-order spin-spin couplings)
\be
r_c^2\equiv r_{\rm EOB}^2+a_0^2\left(1+\dfrac{2M}{r_{\rm EOB}}\right) \; ,
\ee
where, as mentioned before, $r_{\rm EOB}$ is the radial EOB coordinate.
The spin combination
\be
a_0=a_1+a_2
\ee
is the effective Kerr parameter, while $G^{\rm phys}_S(r_{\rm EOB},m_1,m_2,S_1,S_2)$ and 
$G^{\rm phys}_{S_*}(r_{\rm EOB},m_1,m_2,S_1,S_2)$ are the two spin-orbit coupling functions.
In addition, in Eq~\eqref{eq:Horb} we have the radial potential $A(r_{\rm EOB},m_1,m_2,S_1,S_2)$,
the function \hbox{$Q_4\equiv 2\nu(4-3\nu)p_{r_*}^4 u_c^2$} and the radial momentum 
${P_{r^{\rm EOB}_*}\equiv \sqrt{A/B} P_{r^{\rm EOB}}}$ (canonically conjugate to a certain 
tortoise-like radial coordinate $r^{\rm EOB}_{*}$), which introduces the second EOB 
potential function $B(r_{\rm EOB},m_1,m_2,S_1,S_2)$.
In this framework, the dynamics is determined by the structure
of the functions $(A,B,G_{S_*}^{\rm phys},G_{S}^{\rm phys})$, for example
as defined in Refs.~\cite{Damour:2014sva}, and they are chosen such to incorporate
explicitly the (spinning) test-particle limit. Going now to the circular
limit, $P_{r_*^{\rm EOB}}=0$, and defining the dimensionless versions of the
spin-orbit coupling functions $G_S\equiv M^3 G^{\rm phys}_{S}$ and $G_{S_*}\equiv M^3 G_{S_*}^{\rm phys}$,
as well as the dimensionless quantities
$\hat{H}_{\rm eff}\equiv H_{\rm eff}/\mu$ and $\hat{P}_\phi\equiv P_\phi/(M\mu)$,
we have
\be
\hat{H}_{\rm eff}=G_S\hat{P}_\phi S + G_{S_*} \hat{P}_\phi S_*+\sqrt{A\left(1+\hat{P}_\phi^2 U_c^2\right)} \; ,
\ee
where 
\be
\label{eq:defUc2}
U_c^2 \equiv \left(\dfrac{M}{r_c}\right)^2\equiv \dfrac{U^2}{1+\hat{a}_0^2 U^2(1+2 U)} \;  ,
\ee
and $\hat{a}_0\equiv a_0/M$. Here we have defined the inverse EOB radial coordinate
\be
 U\equiv M/r_{\rm EOB}  \equiv  1/\hat{R}\; ,
\ee 
whose relation to the standard Schwarzschild radial coordinate
$\hat{r}\equiv r/M$ will be discussed below in Sec~\ref{sec:HamDyn_RadiusRelations}.
Replacing $(S,S_*)$ with the dimensionless spin variables $(\hat{a}_1,\hat{a}_2)$,
one has
\begin{align}
\hat{H}_{\rm eff}&=\hat{P}_\phi\left\{\left[(X_1)^2\hat{a}_1 + (X_2)^2\hat{a}_2\right]G_S+\nu(\hat{a}_1+\hat{a}_2)G_{S_*}\right\}\nonumber\\
               &+\sqrt{A\left(1+\hat{P}_\phi^2 U_c^2\right)},
\end{align}
where $X_{1,2}=m_{1,2}/(m_1+m_2)$. 
Let us now go to the limit where one mass
is much smaller than the other one, $\mu\equiv m_2\ll M\equiv m_1$, so 
that we have ${\hat{H}=H/\mu=\hat{H}_{\rm eff}}$.
At this stage we still allow $(\hat{a}_1,\hat{a}_2)\neq 0$, 
i.e.\ we keep also the leading-order term (in the mass ratio) proportional to
$\hat{a}_1$, so to consider a {\it spinning} particle on a Kerr background.
Explicitly focusing only on the spin-orbit part, from  $X_1\to 1$ and $X_2\to m_2/m_1$ one gets
\be
\label{eq:HamDyn_SO}
\hat{H}_{\rm SO}\approx \hat{P}_\phi \left\{G_S\left[\hat{a}_1+\left(\dfrac{m_2}{m_1}\right)^2\hat{a}_2\right]
+G_{S_*}\dfrac{m_2}{m_1}\left(\hat{a}_1+\hat{a}_2\right)\right\},
\ee
and keeping only the leading-order terms in $\hat{a}_1$ and $\hat{a}_2$, one finally gets
\begin{align}
\label{eq:HamFinal}
\hat{H}=\hat{P}_\phi\left(G_S\hat{a}_1 + G_{S_*}\dfrac{m_2}{m_1}\hat{a}_2\right)+\sqrt{A\left(1+\hat{P}_\phi^2 U_c^2\right)}
\end{align}
and the functions $G_S$ and $G_{S_*}$ read~\cite{Damour:2014sva,Bini:2015xua},
\begin{align}
\label{eq:GS}
G_S   & = 2 U U_c^2,\\
\label{eq:Gss}
G_{S_*}& = U_c^2\left[\dfrac{r_c\nabla\sqrt{A_{\rm eq}}}{1+\sqrt{Q}}+\dfrac{(1-\nabla r_c)\sqrt{A_{\rm eq}}}{\sqrt{Q}}\right],
\end{align}
where $A_{\rm eq}$ and $B_{\rm eq}$ are the Kerr potentials as defined in Eqs~(10) and~(11) of Ref.~\cite{Damour:2014sva},
and ${\nabla\equiv (B_{\rm eq})^{-1/2}d/dr^{\rm EOB}}$ is the proper radial gradient and ${Q\equiv 1+\hat{P}_\phi^2U_c^2}$.
Note that the $U_c$ entering these two functions is given by Eq~\eqref{eq:defUc2} above, with the Kerr parameter 
$\hat{a}_0$ that is now given, at the same order, by
\be
\hat{a}_0=\hat{a}_1 +\dfrac{m_2}{m_1}\hat{a}_2=\hat{a}_1 + \sigma, 
\ee
where we recall that $\hat{a}_{1,2}=a_{1,2}/m_{1,2}$, Eq~\eqref{eq:defS}, 
and ${\sigma\equiv S_2/(m_1 m_2)=S_2/(M\mu)}$, so that also 
higher-order spin-spin couplings are included in Eq~\eqref{eq:HamFinal}.

Restricting now to the simplest case, the Schwarzschild background, i.e.~$\hat{a}_1=0$,
and keeping only terms {\it linear} in $\sigma$, i.e. $U=U_c$, we obtain the following 
expression for the circular Hamiltonian of a spinning particle on a Schwarzschild background
\be
\label{eq:Hcirc}
\hat{H}=G_{S_*}(U,\hP_\phi)\hP_\phi\sigma + \sqrt{A(U)(1+\hP_\phi^2 U^2)} \; ,
\ee
where the spin-orbit coupling function is still given by Eq~\eqref{eq:Gss},
but with $A_{\rm eq}=1-2U$ and $B=A^{-1}$.
Expression~\eqref{eq:Hcirc} constitutes the central piece
of the Hamiltonian dynamics considered in this work.

Circular orbits are defined from the Hamiltonian in Eq~\eqref{eq:Hcirc} 
in the standard way, demanding
\be
\label{eq:pphi}
\partial_U \hat{H}(U,\hP_\phi;\sigma)=0. 
\ee
For a given $U$, this equation is solved to obtain the corresponding angular 
momentum $\hP_\phi^{\rm circ}(U;\,\sigma)$.
Then, from $\hP_{\phi}^{\rm circ}(U;\,\sigma)$, one obtains
\be
\label{eq:HamOmegaDef}
M\Omega\equiv \partial_{\hP_\phi}\hat{H}(U,\hP_\phi;\sigma)\vert_{\hP_{\phi}^{\rm circ}}\; .
\ee
This is the way we construct the circular dynamics that feeds the Teukolsky
equation and from which we compute the GW energy fluxes (see Sec~\ref{sec:HamDyn_CEOID} below).
In the plots we label this numerically found CEO data by ``HamNum''
in order to distinguish it from the linear in sigma analytical formulas
derived below in Sec~\ref{sec:HAM_linsigma} (labeled by ``HamAna'').

The circular dynamics can be characterized in a gauge invariant way 
in terms of the circular energy, the angular momentum and the frequency
parameter
\be
\label{eq:def_x}
 x\equiv (M\Omega)^{2/3} \; ,
\ee
which also allows to compute $x(U)$ via Eq~\eqref{eq:HamOmegaDef}.
Plugging  $\hP_\phi^{\rm circ}(U;\,\sigma)$
into the Hamiltonian, Eq~\eqref{eq:Hcirc}, one gets $\hat{E}_{\rm circ}(x)$
when using the relation between $U$ and $x$.
In the case of a {\it nonspinning} particle on Schwarzschild, this procedure is fully analytic
and yields the well-known expressions
\begin{align}
\label{eq:E0Schw}
\hat{E}_{0}    &=\dfrac{1-2x}{\sqrt{1-3x}} \; ,\\
\label{eq:j0Schw}
\hP_{\phi}^{0}&=\dfrac{1}{\sqrt{x(1-3x)}} \; .
\end{align}
When $\sigma\neq 0$, it is necessary to solve Eq~\eqref{eq:pphi} numerically so
that $\hat{E}^{\rm circ}(x)$ and $P_{\phi}^{\rm circ}(x)$ are obtained parametrically.
Computing the energy and angular momentum curves along circular orbits, $\hat{E}^{\rm circ}(x,\sigma)$ and 
$\hat{P}_{\phi}^{\rm circ}(x,\sigma)$, or even the relation $\hat{E}^{\rm circ}(P_{\phi}^{\rm circ})$,
is a useful tool to compare  unambiguously the circular Hamiltonian dynamics
with the dynamics obtained from the MP equations, see Sec~\ref{sec:dynamics_energeticsANDisco}.

\subsection{Numerical procedure for CEOs}
\label{sec:HamDyn_CEOID}

The above descriptions might appear a bit complicated due to the
choice to give a general discussion but in fact it is
easy to compute the CEO data. The starting point is Eq~\eqref{eq:Hcirc},
where we simply fix the EOB coordinates $(\hat{R},\phi)$ 
and the spin parameters $(\hat{a}_1,\sigma)$ as desired.
It remains to find data for $\Omega$ and $P_\phi$.
To find the data for $P_\phi$,
we differentiate $\hat{H}$ with respect to $U$ and demand that the resulting expression vanish, Eq~\eqref{eq:pphi}.
In that expression we insert the orbital distance $\hat{R}$ and spin $\sigma$ to make the right-hand-side a function of $P_\phi$ only.
This gives a condition to fix $P_\phi$, which we solve numerically.
Finally, $\partial_t \phi \equiv \Omega$ is obtained
by differentiating $\hat{H}$ with respect to $P_\phi$
and inserting the values for $\hat{R}$ and $P_\phi$.
Thus, CEOs are given by the roots of two algebraic equations,
which we do using MATLAB's \texttt{fzero} routine.
Before proceeding note that a given EOB radial coordinate is not trivially linked
to a corresponding BL radial coordinate, see Sec~\ref{sec:HamDyn_RadiusRelations} and Sec~\ref{sec:HamDyn_TEapplication}.

\subsection{Linear-in-$\sigma$ energy and angular momentum}
\label{sec:HAM_linsigma}

It is possible to obtain fully analytic expressions of the energy and angular
momentum along circular orbits {\it if} we work consistently at linear order in $\sigma$.
Such result will be useful in various respects, e.g., in driving a comparison with the MP energetics and
the HamNum energetics and to cross-check with other literature results obtained analytically
from the MP at the quadrupolar order.

So, working now for a while at linear order in the spin, to determine $\hP_\phi^{\rm circ}(U;\,\sigma)$,
we pose  ${\hP_\phi^{\rm circ}(U;\,\sigma)=\hP_\varphi^0(U)+\sigma \hP^1_\phi(U)}$, where
$\hP_\phi^0$ is the solution of $\partial_U\hat{H}(U,\hP_\phi;0)=0$
(i.e., Eq~\eqref{eq:j0Schw} above with $x=U$)
and we just need to solve for $\hP^1_\phi$.
From the definition of the orbital frequency
$M\Omega\equiv \partial_{\hP_\phi}\hat{H}(U,\hP_\phi;\sigma)$ we can obtain $x$ and
the link between the inverse radial coordinate $U$ and $x$ along circular orbits
\be
\label{eq:Uvsx}
U(x;\,\sigma)\equiv x+\dfrac{x^{5/2}}{1+\sqrt{\dfrac{1-2x}{1-3x}}}\sigma \; .
\ee
Using this relation, one finally finds the linear in $\sigma$ expression of the angular momentum
along circular orbits
$\hP_\phi^{\rm circ}(x;\,\sigma)\equiv \hP_\phi^{\rm circ}(U(x;\sigma);\,\sigma)$,
which reads explicitly
\begin{align}
\label{eq:PphiCirc}
\hP_\phi^{\rm circ}(x;\,\sigma)=\dfrac{1}{\sqrt{x(1-3x)}}-\left(1-\dfrac{1-4x}{\sqrt{1-3x}}\right)\sigma \ ,
\end{align}
and, by inserting the two equations above into Eq~\eqref{eq:Hcirc}, the corresponding
expression for the energy reads
\begin{align}
\label{eq:ecirc}
\hat{E}^{\rm circ}(x;\,\sigma) = \dfrac{1-2x}{\sqrt{1-3x}}- \dfrac{x^{5/2}}{\sqrt{1-3x}}\sigma \; ,
\end{align}
which manifestly contains the nonspinning particle limit, see Eq~\eqref{eq:E0Schw}.
These two relations fully characterize the circular orbits of a spinning particle
on a Schwarzschild BH at linear order in the spin. We will use these 
in Sec~\ref{sec:dynamics_energeticsANDisco} to drive comparisons with the 
corresponding energetics obtained numerically from the Hamiltonian
as well as with the energetics from the MP equations.

\subsection{The spin-induced ISCO shift}
\label{sec:ISCO_spinning}

The ISCO location and the corresponding minimal angular momentum 
is defined by the two equations
\begin{align}
\partial_U\hat{H}(U_{\rm ISCO},\hP_\phi^{\rm ISCO};\sigma)&=0 \nonumber \\
\partial_U^2\hat{H}(U_{\rm ISCO},\hP_\phi^{\rm ISCO};\sigma)&=0.
\label{eq:ISCOs_spinningparticle_Hamiltonian}
\end{align}
This system can be solved numerically to obtain $(U_{\rm ISCO},\hP_{\phi}^{\rm ISCO})$
and eventually this yields the ISCO frequency parameter $x_{\rm ISCO}^{\rm HamNum}$ 
that we list in the sixth column of Table~\ref{tab:ISCOs_x}.

As we did above, it is also instructive to solve this system analytically
working at liner order in $\sigma$. We obtain
\begin{align}
\label{eq:lsoU}
U^{\rm ISCO}&=\dfrac{1}{6}+\sigma \dfrac{1}{18\sqrt{2}} \ ,\\
\label{eq:lsoL}
\hat{P}_{\phi}^{\rm ISCO}&=\pm 2\sqrt{3}+\sigma\left(\dfrac{\sqrt{2}}{3}-1\right) .
\end{align}
Using the inversion of the link between $x$ and $U$ in Eq~\eqref{eq:Uvsx},
\be
\label{eq:xvsU}
x(U)=U-\dfrac{U^{5/2}}{1+\sqrt{\dfrac{1-2U}{1-3U}}}\sigma,
\ee
we also obtain 
\begin{align}
\label{eq:xISCO}
x^{\rm ISCO}=\dfrac{1}{6} + \dfrac{\sigma}{12\sqrt{6}} 
\end{align}
for the ISCO frequency parameter.
Note that below we will write Eq~\eqref{eq:xISCO} as $x^{\rm ISCO}_{\rm HamAna}$ to
distinguish it from the numerical solution of Eqs~\eqref{eq:ISCOs_spinningparticle_Hamiltonian}.
The ISCO energy is
\be
\label{eq:lsoE}
\hat{E}^{\rm ISCO}= \dfrac{2\sqrt{2}}{3}-\dfrac{\sqrt{3}}{108}\sigma \qquad .
\ee
These results, more concretely Eqs~\eqref{eq:lsoL}, \eqref{eq:xISCO} 
and~\eqref{eq:lsoE}, coincide with the corresponding linear-in-spin expressions 
obtained analytically by Bini, Faye and Geralico~\cite{Bini:2015zya} 
starting directly from the MP. For completeness, we quote also their
expression for $x_{\rm ISCO}$, which also includes the $\sigma^2$ term and reads
\begin{align}
\label{eq:lsoBFG}
x_{\rm ISCO}^{\rm BFG}=\dfrac{1}{6}+\dfrac{\sigma}{12\sqrt{6}}+\dfrac{\sigma^2}{216}.
\end{align}
Thus, our linear in $\sigma$ calculation gives a useful consistency check that the circular 
dynamics of Ref.~\cite{Bini:2015zya} is precisely the same as the one provided by 
the  limit of the EOB Hamiltonian. This finding will turn out useful
below to obtain a linear in $\sigma$ link between the EOB and the BL
radial coordinates along circular orbits.

\subsection{Relation between the EOB and Schwarzschild radial coordinates}
\label{sec:HamDyn_RadiusRelations}

For our application of the dynamics to the Teukolsky equation
it is necessary to have an explicit connection between
the EOB and the Schwarzschild radial coordinates, or equivalently their inverses
$U=M/r^{\rm EOB}$ and $u=M/r$, at least at linear order in $\sigma$.
For a {\it nonspinning} particle, we would just have $u=U$.
For a {\it spinning} particle, however, this relation will be corrected by
a term linear in sigma. This relation does not seem to exist in the 
literature. Here we shall derive it in the simplifying case of 
circular orbits.

An easy way to relate $u$ with $U$ is to equate the two functions
$x(u)$ and $x(U)$, as obtained
from the MP using a given SSC on the one hand and from the Hamiltonian on the other hand. 
The relation $x(U)$ was already given in Eq~\eqref{eq:xvsU}.
Similarly, from the MP with the T SSC, 
one obtains at linear order in $\sigma$,
\be
\label{eq:x_u}
x(u) =  u -  u^{\frac{5}{2}} \sigma,
\ee
which follows when linearizing Eq~(22) of Paper~\RM{1} 
or Eq~(4.26) of Ref.~\cite{Tanaka:1996ht}. Equating the
two expressions for the frequency parameter, we thus find
\begin{align}
 \label{eq:UEOB_to_uBL_linearinsigma}
 u = U + \dfrac{\sqrt{\dfrac{1-2 U}{1-3 U}}}{1+\sqrt{\dfrac{1-2 U}{1-3 U}}}\, U^{5/2}\sigma \; ,
\end{align}
and correspondingly
\begin{align}
 \label{eq:rEOB_to_rBL_linearinsigma}
\hat{r} = \hat{R} - \dfrac{1}{\hat{R}\sqrt{1-\dfrac{3}{\hat{R}}}}\sigma .
\end{align}
Of course, these results can also be applied to
determine the ISCO shift of the BL radius. 
From Eq~\eqref{eq:lsoU} one obtains the $\sigma$-dependent ISCO shift in EOB coordinates
\begin{align}
 \hat{R}^{\rm ISCO}=6-\sqrt{2} \sigma ,
\end{align}
and from Eq~\eqref{eq:rEOB_to_rBL_linearinsigma} we get
\begin{align}
 \hat{r}^{\rm ISCO}=6-\dfrac{2}{3}\sqrt{6}\,\sigma,
\end{align}
consistently with the corresponding correction in Ref.~\cite{Bini:2015zya},
see Eq~(4.38)-(4.40) there.

\subsection{Orbital dynamics for the Teukolsky Equation source term}
\label{sec:HamDyn_TEapplication}

The particle source term of the Teukolsky equation is assembled from the variables~\eqref{eq:spinning_particle_vars}.
Thus we cannot directly use the natural variables of the EOB Hamiltonian formalism 
but have to further process these.
First, we calculate the BL-coordinates and their time derivatives from the EOB ones.
While the angular coordinates are the same, the BL-radius is computed from the
ansatz made in Sec~\ref{sec:HamDyn_RadiusRelations},
i.e.\ at linear order in $\sigma$ we assume $x(u)=x(U)$. As a matter of choice,
in the code the resulting equation
\be
u-u^{5/2}\sigma = U - \dfrac{U^{5/2}}{1+\sqrt{\dfrac{1-2U}{1-3U}}}\sigma
\ee
has been solved numerically for $u$, instead of using
the strict analytical linear in $\sigma$ solution, Eq~\eqref{eq:rEOB_to_rBL_linearinsigma}.
We have checked that the results are essentially the same;
e.g.\ for $\sigma=0.9$ and over the interval $R\in(5,30)$ the solutions
for $r$ differ at most by $0.5\%$.
Due to the linear in $\sigma$ approximations used,
there is anyway a small uncertainty in the interpretation of
the Hamiltonian CEO dynamics in terms of $r$.
That is but one reason why we prefer to discuss the results in terms of gauge
invariant parameters like the frequency whenever possible.
Continuing to process the EOB variables, the tangent vector can be computed 
from the coordinate velocities using
\begin{align}
 v^i = \frac{d X^i}{dt} v^t \; ,
\end{align}
where we compute $v^t$ according to Eq~\eqref{eq:lapse}. To compute $p_\mu$ and $S^{\mu\nu}$,
we need to impose again a SSC. For simplicity we choose here the linearized T SSC, Eq~\eqref{eq:T_SSC}.
Recall that at linear order the T SSC is equivalent to the P SSC and we have 
\begin{align}
 p_\mu = \mu v_\mu \; ,
\end{align}
which directly relates the kinematical momenta to the four velocity.
The spin tensor $S^{\mu\nu}$ is computed according to Eq~\eqref{eq:SpinTensEQ}, with $V^\mu=v^\mu$.
The time derivatives are computed numerically, but for CEOs found to vanish anyway.
As a side remark, note that the EOB momenta that are evolved by the Hamiltonian EOM are
not used at all to compute the variables~\eqref{eq:spinning_particle_vars}.

\section{Energetics and the ISCO shift}
\label{sec:dynamics_energeticsANDisco}

In the next section we want to analyze the different circular
dynamics that we have produced using the respective approaches described above.
We discuss the energetics and how the shift of the ISCO
due to the particle's spin changes between the various cases.
In the analysis of the different prescriptions we pick the Hamiltonian case as
the reference solution. This does not imply the Hamiltonian description would be more
correct than the others, it is only motivated by our research agenda, i.e.\
we plan to use our findings for the EOB Hamiltonian case to model waveforms.

\subsection{Energetics of circular orbits}

\begin{figure*}[t]
  \centering  
  \includegraphics[width=0.45\textwidth]{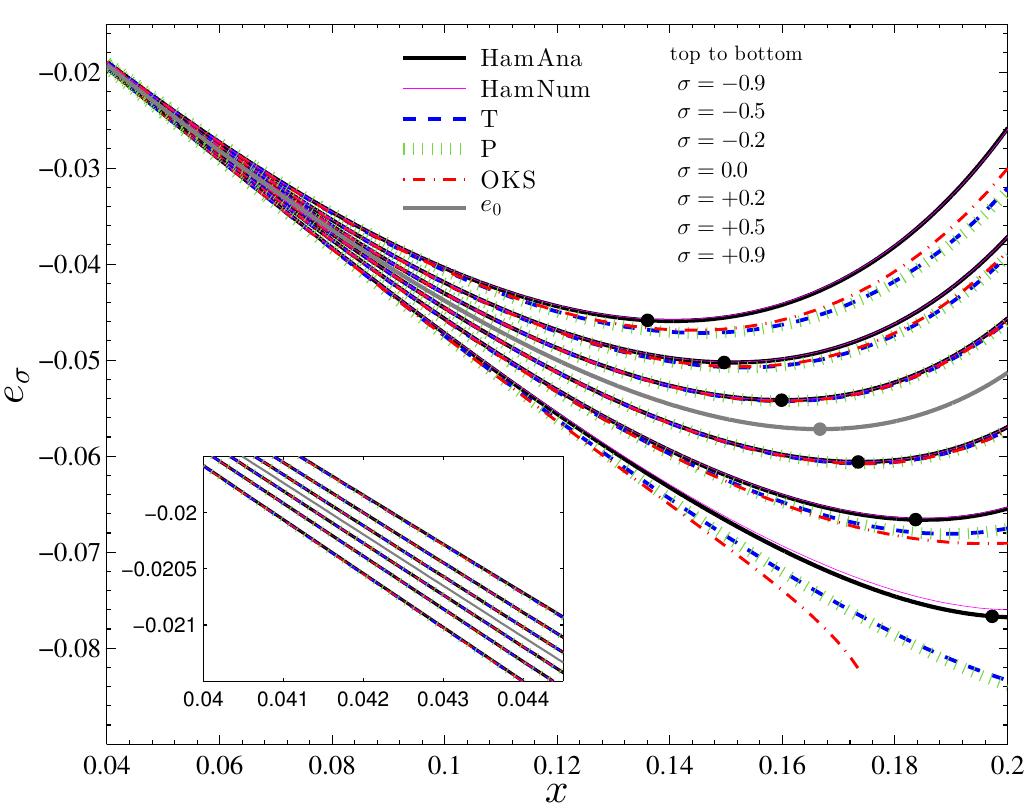} 
  \hfil
  \includegraphics[width=0.45\textwidth]{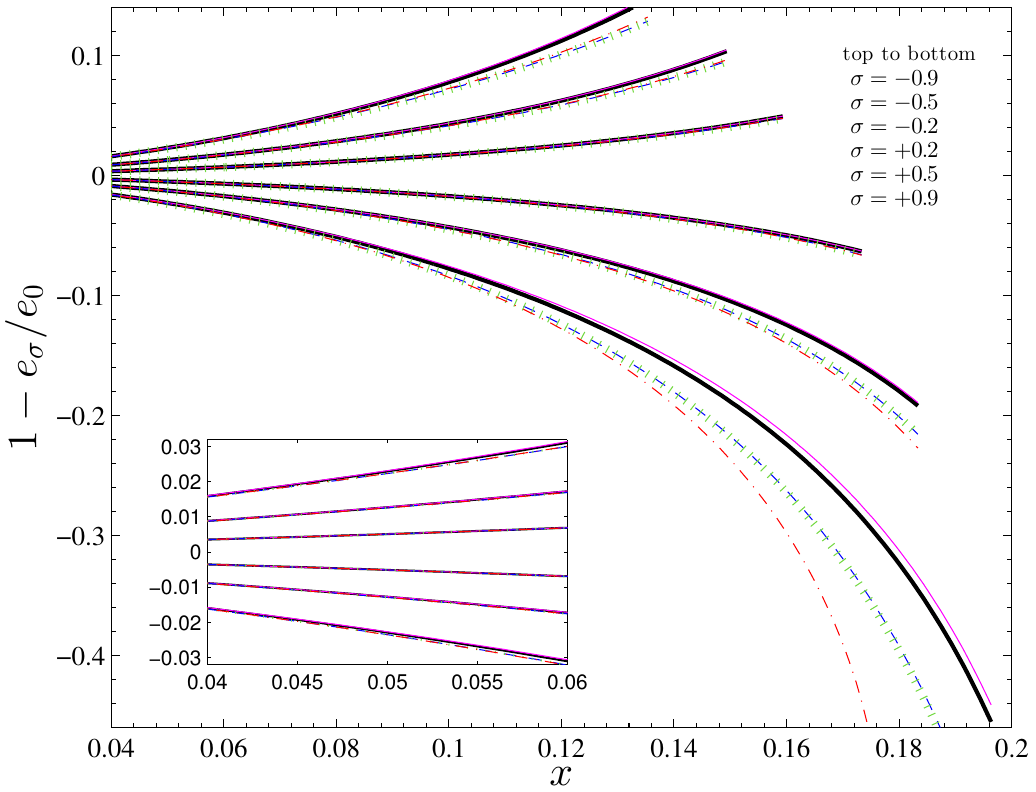} 
  \\
  \caption{Analysis of spin effects on the energetics of circular orbits.
    Left panel:
    Contrasting the binding energy $e_{\sigma}(x)\equiv \hat{E}_{\sigma}(x)-1$ for different 
    values of the particle spin $\sigma$ as obtained on the one hand analytically (solid black) at linear order in the spin
    from Hamiltonian dynamics, i.e.\ using $\hat{E}^{\rm circ}_{\sigma}(x)$ of Eq~\eqref{eq:ecirc},
    and on the other hand numerically from the Hamiltonian (solid magenta), from
    the MP with the T SSC (blue dashed), the P SSC (green dotted) and the OKS SSC (red dash-dotted) respectively.
    The filled black markers are the analytic energy values at the analytic ISCO frequencies, obtained from
    $\hat{E}^{\rm circ}_{\sigma}(x_\textrm{ISCO}^{\rm HamAna})$.
    Right panel:
    The (fractional) spin contributions (see Eq~\eqref{eq:eSpin}), shown 
    up to $x_{\rm ISCO}^{\rm HamAna}$ (see Eq~\eqref{eq:xISCO}).
  } 
  \label{fig:Ex}
\end{figure*}

\begin{figure*}[t]
  \centering  
  \includegraphics[width=0.32\textwidth]{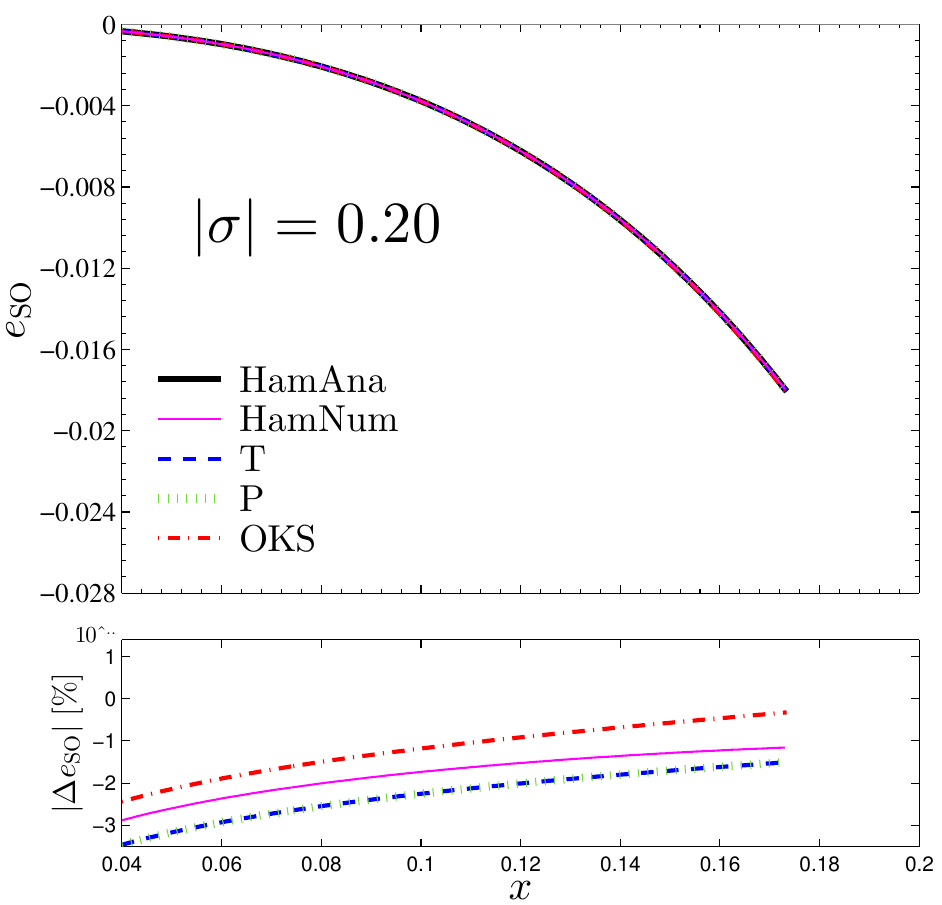}
  \includegraphics[width=0.32\textwidth]{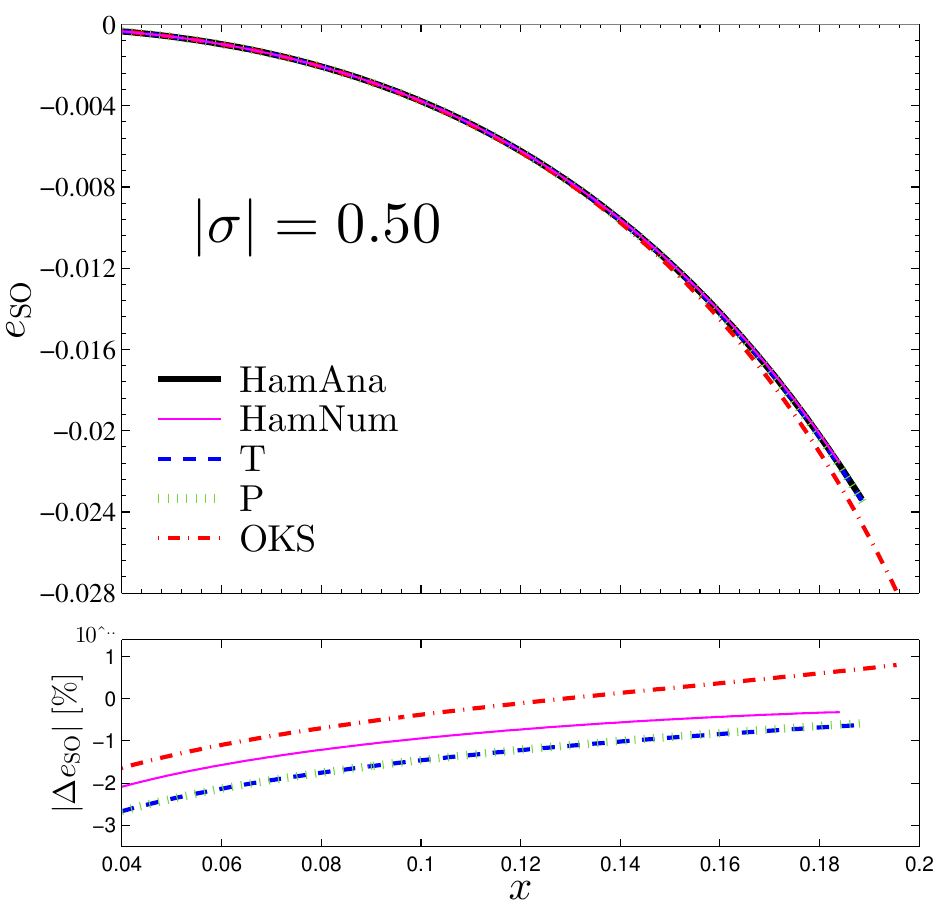}
  \includegraphics[width=0.32\textwidth]{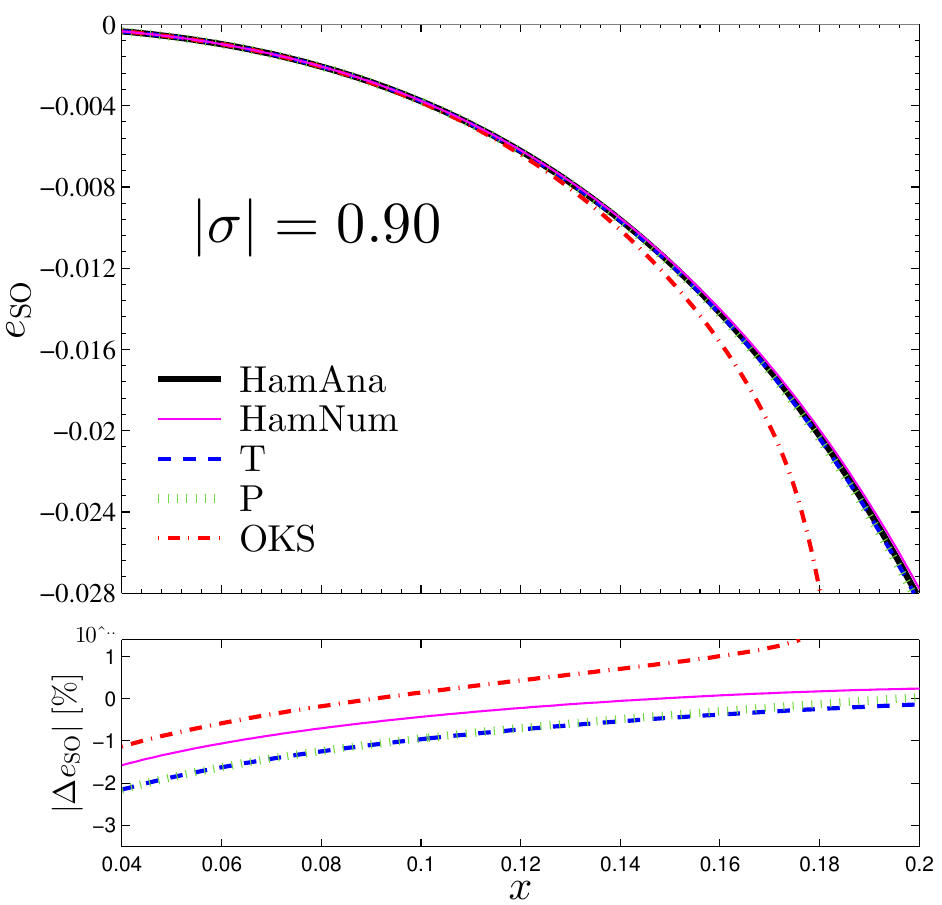}
  \caption{
    Analysis of the spin-orbit contribution, $e_\textrm{SO}$,
    to the energetics of circular orbits.
    Top panels: 
    Comparing the analytic formula for the spin-orbit contribution in the
    Hamiltonian case (HamAna), as read-off from Eq~\eqref{eq:ecirc},
    with $e_\textrm{SO}(x,\sigma)$ for the different dynamical prescriptions,
    computed using Eq~\eqref{eq:eSO}.
    From left to right the panels show three representative values
    of the spin: $|\sigma|=0.2,0.5,0.9$.
    Bottom panels: 
    Relative differences 
    of the SSC curves with respect to the Hamiltonian curves,
    $| \Delta e_\textrm{SO} | = | 1 - e_\textrm{SO}^{\rm X} / e_\textrm{SO}^{\rm HamAna} |$,
    where X refers to the HamNum, T, P, and OKS case respectively.
    Note that the vertical scale is logarithmic here.    
   }  
  \label{fig:eSO}
\end{figure*}

To drive gauge invariant comparisons of the dynamics,
one may analyse the binding energy as a function of
the angular momentum, as e.g.\ in ~\cite{Steinhoff:2012rw}.
Alternatively, as done here, one can
consider the binding energy as a function of the orbital frequency $\Omega$, or 
of the frequency parameter $x$.
Note that at a given BL $r$ the orbital frequency $\Omega$
differs between the respective dynamics, and thus also $x$ does.

The energetics are represented
via the reduced binding energy function 
\be 
e_{\sigma}(x) := \hat{E}(x,\sigma)-1 \ ,
\ee 
where $\hat{E}(x,\sigma)$ is the mass-reduced energy constant on
a circular orbit, see Eqs~\eqref{eq:ConEq_E},~\eqref{eq:ecirc}, and~\eqref{eq:dimless_quantities} respectively.
From $e_{\sigma}(x)$ we seek to isolate 
i)~the contribution due to the body's spin and, in particular,
ii)~the spin-orbit (SO) contribution
\footnote{A similar analysis has
  been applied to study the spin-orbit interaction in numerical
  relativity simulations of neutron star mergers
  \cite{Bernuzzi:2013rza}.}.
For i) we simply compare the energetics with the nonspinning particle limit,
i.e.\ we consider the fractional differences
\be
\label{eq:eSpin}
1-\frac{e_{\sigma}}{e_{0}} \; .
\ee
For ii) we note that from the SO Hamiltonian Eq~\eqref{eq:HamDyn_SO} an ansatz for the energy 
function can be motivated in the Newtonian limit, i.e.\ at low frequencies,
which reads 
\be
\label{eq:Exsigma}
\hat{E}(x,\sigma) = \hat{E}_{0}(x) + h_\textrm{SO}^{\rm lin}(x)\sigma + \O(\sigma^2) \ ,
\ee
where the first term corresponds to the nonspinning dynamics, and the
second term to the SO interaction at linear order in
$\sigma$. The $\O(\sigma^2)$ terms describe either high-order SO
contributions or spin-spin self interactions. 
By construction, the Hamiltonian dynamics does not contain 
$\O(\sigma^2)$ terms, so $h_\textrm{SO}^{\rm lin}(x)$ is
analytical and can be read off from Eq~\eqref{eq:ecirc}. From the MP
dynamics we extract the SO contribution according to
\be
\label{eq:eSO}
e_\textrm{SO}(x,|\sigma|) := \frac{1}{2 \sigma} \left( \hat{E}(x,+|\sigma|) - \hat{E}(x,-|\sigma|) \right) \ .
\ee
This formula is obviously insensitive to the sign of the spin because
it employs pairs of datasets with spins of the same
magnitude $|\sigma|$ but opposite in sign.  
Note that $e_\textrm{SO}(x,\sigma)$ computed as above  
depends on $\sigma$ if the \textit{additive} ansatz
\eqref{eq:Exsigma} does not hold, i.e. presumably at high-frequencies and at 
high spins. Thus, in general we expect that 
$e_\textrm{SO}=e_\textrm{SO}(x,\sigma)$, while for fully linear-in-spin
expressions like the HamAna formula~\eqref{eq:ecirc}
the quantity $e_\textrm{SO}$ is actually spin-independent, ${e_\textrm{SO}=e_\textrm{SO}(x,0)}$.

Let us now contrast the just introduced quantities for the different dynamics.
First, the full energy curves $e_{\sigma}(x)$ are shown 
for seven representative values of the spin $\sigma \in (0,\pm 0.2 , \pm 0.5, \pm 0.9)$
in the left panel of Fig~\ref{fig:Ex},
illustrating the system's energetics along circular orbits from large distances
(low frequencies) to small distances (high frequencies) close to the ISCO,
and in some cases even beyond.
The black dot markers refer to the value of $e_{\sigma}(x)$ at $x=x_\textrm{ISCO}$
as computed by Eq~\eqref{eq:xISCO}.
The figure includes lines for the T SSC, the P SSC and the OKS SSC, as well as for
the Hamiltonian formalism. 
Note that the Hamiltonian formalism actually provides two results,
namely, the analytic approximation (``HamAna''), Eq~\eqref{eq:ecirc}, and 
the full numerical solution of Eqs~\eqref{eq:pphi} and~\eqref{eq:HamOmegaDef} inserted
into Eq~\eqref{eq:Hcirc} (``HamNum''). 
The comparison of these two is an important
corollary result for understanding the character of the EOB Hamiltonian though this
is not explored in detail here. 

Looking at the panels, 
at first sight the energetics of the three MP dynamics are qualitatively comparable
with each other and with the Hamiltonian counterpart.
While at small $x$ the curves are visually on top of each other (see inset),
one clearly observes that the spin interactions become more significant as $x$ 
increases (smaller orbital radii) \textit{and} as
$|\sigma|$ increases. Towards the ISCOs the analytic formula for the
energy, Eq~\eqref{eq:ecirc}, shows significant differences with respect to the one from the MP dynamics.
In general, in the regime of large frequencies, $x\gtrsim 0.1$,
positive spins, i.e.\ spins aligned with the orbital angular momentum,
involve the worst mutual agreement between the curves;
among the various cases the OKS SSC shows
the least consistency with the Hamiltonian reference case for large positive spins.
By contrast, the numerical solution of the Hamiltonian dynamics is in visual agreement
with the analytical approximation over the whole spin and frequency range considered.
Interestingly, also the T SSC and the P SSC are mutually consistent.

Let us now compare the energetics with the {\it nonspinning} particle limit.
The right panel of Fig~\ref{fig:Ex} shows the fractional
spin contribution, i.e.\ Eq~\eqref{eq:eSpin},
up to $x_{\rm ISCO}^{\rm HamAna}$.
For small spin magnitudes $\sigma=\pm0.2$ all dynamical prescriptions are
in perfect agreement over the whole frequency range considered.
The relative differences with respect to the nonspinning limit are small,
namely $\lesssim5$\%.
For larger spin magnitudes the energetics deviate, of course,
much more from the nonspinning limit; in general, we observe
that at a given spin magnitude positive
spins entail a larger deviation from the nonspinning limit
than negative spins; e.g.\ for $|\sigma|=0.9$ the differences close to the ISCO
are $\sim 50\%$ for $\sigma=0.9$ while they are $\sim 15 \%$ for $\sigma=-0.9$.
For large spin magnitudes the energetics of the different prescriptions
are still in agreement at small frequencies (see inset) but
show again significant variations at large frequencies, $x \gtrsim 0.1$.
Furthermore, note that this plot highlights the repulsive character of the spin 
interactions for $\sigma>0$ (spin aligned to orbital angular momentum)
and the attractive character for $\sigma<0$ (spin anti-aligned to
orbital angular momentum). 

Next, the SO contribution is analyzed in  
Fig~\ref{fig:eSO}, which shows $e_\textrm{SO}(x,\sigma)$ as computed by Eq~\eqref{eq:eSO} for the
three MP dynamics (T, P, OKS) and the numerical solution of the Hamiltonian dynamics (HamNum).
These curves are contrasted with the analytical term $e_\textrm{SO}^{\rm HamAna}=h_\textrm{SO}^{\rm lin}$,
as read off from Eq~\eqref{eq:ecirc} (HamAna).
From left to right the panels of Fig~\ref{fig:eSO} refer to the 
representative spin values $|\sigma| \in (0.2,0.5,0.9)$.
The panels show that the SO interaction
is modeled in a qualitatively compatible way in all the different cases;
only for large frequencies and for $|\sigma|=0.9$ the OKS SSC
entails a drastic inconsistency in the SO part compared with the other approaches.
Beholding the figure, it is striking in all three panels that the
P SSC and the T SSC behave visually equivalently in mutual comparison.
The bottom panels show the relative differences 
with respect to the linear in sigma analytic formula,
$| \Delta e_\textrm{SO} | = |  1 - e_\textrm{SO}^{\rm X} / h_\textrm{SO}^{\rm lin} |$.
The relative differences of the P and the T SSC to the Hamiltonian dynamics
grow from only 0.001\% (at $x\sim0.04$) to 0.01\%
(at $x\sim x_\text{ISCO}$) for $|\sigma|=0.2$.
For $|\sigma|=0.9$ the differences increase by approximately
one order of magnitude, which amounts to a difference of $\lesssim 1 \%$ close to the ISCO.
These differences are likely due to nonlinear-in-spin terms included in the MP dynamics
when using the T and P SSCs, which are
invalidating the linear in spin ansatz of the Hamiltonian approach, see Eq~\eqref{eq:Exsigma}. 
The SO interaction observed for the MP OKS dynamics shows instead more significant
differences with respect to both the T and the P SSC as well as to the Hamiltonian reference.
Focusing on small spins, $|\sigma|\sim0.2$, and ``large'' 
separations $x\sim0.04$, the OKS case already manifests a SO contribution rather inconsistent
with the others. The relative differences to the Hamiltonian case are in general one order larger
than those observed for the T and the P SSC. At $|\sigma|=0.9$ the deviations become visually apparent,
in particular at high frequencies the divergence becomes most prominent.
This disagreement is at first sight surprising from the point of view that the OKS SSC
EOM are linear-in-spin as is the HamAna expression. 
However, though the EOM are linear in spin for the OKS SSC,
its energy dependency on the spin appears to be not necessarily linear in spin,
which explains why the SO part can be 
i)~spin-dependent, and
ii)~different from the Hamiltonian analytical formulas.
Furthermore, a disagreement with the Hamiltonian case is reasonable
because the Hamiltonian comes from a by hand
linearization of the T/P MP in spin, whereas the OKS SSC stems from the idea of
choosing an observer such that the MP are linear in spin. Thus, the
Hamiltonian is dynamically closer related to the T/P MP while the OKS formalism comes from
another, independent approach.

\subsection{ISCO Results}

Besides comparing the energetics, the differences in the dynamics
can be analyzed in terms of the spin dependent shift of the ISCO location;
see Appendix~\ref{sec:ISCO_nonspinning} for
a reminder on the notion of the ISCO for a {\it nonspinning} particle.
We report in Tab~\ref{tab:ISCOs_x} the values of $x_{\rm ISCO}$ 
for the different dynamical prescriptions and the different
approximations: the MP with the T SSC, the P SSC and the OKS SSC,  
and the Hamiltonian dynamics, either in form of the analytic linear-in-spin expression
Eq~\eqref{eq:xISCO}, or in the full numerical solution of
Eqs~\eqref{eq:ISCOs_spinningparticle_Hamiltonian}. Additionally, we include
the quadrupolar result as derived by Bini, Faye and Geralico in~\cite{Bini:2015zya} (BFG hereafter). 
We stress that, for the OKS SSC, ISCOs could not be found at spins
$\sigma\gtrsim 0.52$. Similarly the ISCO frequency diverges for the P SSC
at spins $\sigma\gtrsim 0.94$. Among the considered MP options, only the T SSC gives well defined
ISCOs for every $\sigma\in[-1,+1]$ (and, in fact, even beyond that interval).

Browsing through Tab~\ref{tab:ISCOs_x}, one observes that the T and the P SSC
behave similarly; the ISCO values for the frequency parameter agree
in all cases in the first two significant digits, except for the highest spin value $\sigma=0.9$.
A bit worse of an agreement is seen between the numerical Hamiltonian approach (HamNum)
and the linear in sigma analytic formula (HamAna). While the ISCO frequencies
also share the same first two significant digits for small spins $|\sigma| \leq 0.3$
they deviate the stronger the larger is the spin magnitude.
These consistencies between the T and the P SSC on the one hand and HamNum and HamAna on the
other hand became already apparent in the
analysis of the energetics made in the previous section and are confirmed here.
Surprisingly, we also observe strong agreement (2 digits) between $x_{\rm{ISCO}}^{\rm HamNum}$
and $x_{\rm{ISCO}}^{\rm BFG}$, i.e.\ the quadrupolar result of~\cite{Bini:2015zya}.
This result suggests that the Hamiltonian implicitly models higher order spin
terms connected to the spin-induced quadrupole of the body.

\begin{figure}[t]
  \centering  
  \includegraphics[width=0.44\textwidth]{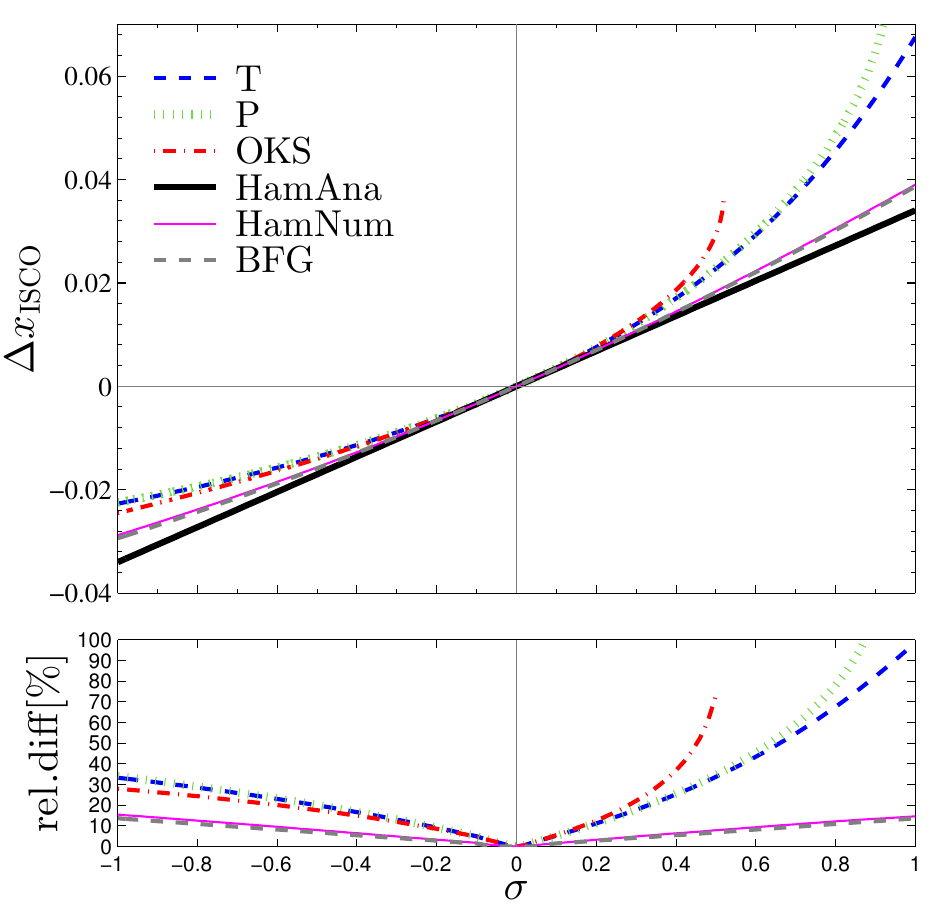} 
  \caption{Top panel: Shift of the ISCO frequency parameter due to the particle's spin,
    $ { \Delta x_{\rm{ISCO}} = x_{\rm{ISCO}} - 1/6 = (M\Omega_{\rm{ISCO}})^{\frac{2}{3}} - 1/6 }$.
    Bottom panel: Relative differences of the ISCO shifts $\Delta x_{\rm{ISCO}}$
    with respect to the linear-in-sigma analytical formula of the Hamiltonian dynamics (HamAna).    
    For the OKS SSC (red, dash-dotted) the ISCO computations fail for spins $\sigma >0.52$ (see discussion in text);
    for the P SSC (green, dashed) for spins $\sigma > 0.94$.
    In general, the figure illustrates how all prescriptions converge
    to the same ISCO frequency as $\sigma \rightarrow 0$.     
    At large values of $\sigma$ the ISCO frequencies are drastically different
    between the various dynamical formalisms.
    In particular, we observe that the ISCOs obtained from the Hamiltonian
    analytically (HamAna) and fully numerically (HamNum) agree well
    with one another ($\lesssim 15\%$), as well as with the quadrupolar
    result of~\cite{Bini:2015zya} (BFG, gray dashed).
    For spins $\sigma<0.2$ also the ISCO frequencies of the T SSC, the P SSC 
    and the OKS SSC are very compatible with one another. }    
  \label{fig:xISCO_shift}
\end{figure}

Table~\ref{tab:ISCOs_x} is complemented by Fig~\ref{fig:xISCO_shift},
which shows in the top panel the ISCO shift 
\begin{align}
\Delta x_{\rm{ISCO}}(\sigma) 
&= x_{\rm{ISCO}}(\sigma) - x_{\rm{ISCO}}(0) \nonumber\\
&= x_{\rm{ISCO}}(\sigma) - \frac{1}{6}  
\end{align}
due to the particle's spin. 
For small spins $|\sigma|\lesssim0.2$ all the different cases are in agreement 
with one another, as expected.
The agreement between the T and the P SSC is striking
over the whole range of $\sigma$, though it deteriorates a bit for large positive spins
where the ISCO for the P SSC is divergent. For negative $\sigma$ also the OKS SSC seems 
compatible with the two. 
For large positive spin magnitudes the plot clearly shows the divergence of the ISCO
within the OKS and the P SSC.
Confirming the impression of Table~\ref{tab:ISCOs_x}, the numerically found ISCOs
of the Hamiltonian dynamics are very close to
the BFG formula, which includes the quadrupolar contributions, over the whole spin range.
This is an interesting but rather
surprising numerical coincidence, with $\sigma^{2}$ terms that are effectively present
in the numerical solution. 
The bottom panel shows the differences with respect to the linear-in-sigma analytical
formula~\eqref{eq:xISCO}. 
For moderate spins $|\sigma|\lesssim0.2$ the ISCO frequencies of the different dynamics
differ by at most $\sim20$\% from the linear in sigma Hamiltonian expression.
As expected, the numerical and the analytical Hamiltonian solution
for the ISCO are quite compatible, with only $\sim 15\%$ deviations at maximum.
Also the BFG result agrees with the HamAna result at that level.

As a side note we mention that, when taking the viewpoint that a SSC is just a gauge choice within
the pole-dipole approximation, one would not expect such
discrepancies between the curves of the three SSCs in Fig~\ref{fig:xISCO_shift}
because $x_{\rm{ISCO}}$ is a gauge invariant quantity.
One can argue that our plots are missing parallel shifts of the curves
along the $\sigma$ axis, since the reference point according to which we define
$\sigma$ depends on the SSC, but even such shifts could not make the curves match each other.
Thus, our results indicate that the different choices of a centroid made in the different
dynamical approaches of the pole-dipole approximation actually lead to
different physical descriptions, rather than only artificial gauge effects. 
Note, however, that if all the multipole moments of the test body were present,
i.e.\ we did not have just a pole-dipole approximation, then a SSC would be just a gauge
choice, see~\cite{Costa:2015} for more details on this issue.

\begin{table}[t]
\caption{Frequency parameter $x$ at the ISCO of a spinning particle on
  a Schwarzschild background computed with different dynamics.
  Entries with backslash $/$ mean that the
  ISCO values for these configurations could not be found, see text.}
\centering
\begin{ruledtabular}
  \begin{tabular}[t]{c c c c c c c c} 
  $\sigma$   &
  $x_{\rm ISCO}^{\rm T }$  &
  $x_{\rm ISCO}^{\rm P }$  &  
  $x_{\rm ISCO}^{\rm OKS }$  &
  $x_{\rm ISCO}^{\rm HamAna}$    &
  $x_{\rm ISCO}^{\rm HamNum}$    &
  $x_{\rm ISCO}^{\rm BFG}$ \\
  \hline
  \hline
  $ 0.90  $ &  0.222448  &  0.231253  &  $/$  &  0.197290  &  0.201378  &  0.201040  \\ 
  $ 0.70  $ &  0.203471  &  0.204538  &  $/$  &  0.190480  &  0.193037  &  0.192750  \\ 
  $ 0.50  $ &  0.189398  &  0.189541  &  0.196051  &  0.183680  &  0.185012  &  0.184830  \\ 
  $ 0.30  $ &  0.178680  &  0.178692  &  0.179119  &  0.176870  &  0.177361  &  0.177290  \\ 
  $ 0.10  $ &  0.170248  &  0.170248  &  0.170257  &  0.170070  &  0.170123  &  0.170120  \\ 
  $ -0.10  $ &  0.163426  &  0.163426  &  0.163420  &  0.163260  &  0.163319  &  0.163310  \\ 
  $ -0.30  $ &  0.157786  &  0.157790  &  0.157670  &  0.156460  &  0.156951  &  0.156880  \\ 
  $ -0.50  $ &  0.153042  &  0.153066  &  0.152642  &  0.149660  &  0.151010  &  0.150810  \\ 
  $ -0.70  $ &  0.149000  &  0.149076  &  0.148144  &  0.142850  &  0.145476  &  0.145120  \\ 
  $ -0.90  $ &  0.145521  &  0.145696  &  0.144062  &  0.136050  &  0.140326  &  0.139800  \\ 
  \end{tabular} 
 \end{ruledtabular}
\label{tab:ISCOs_x}
\end{table}

\section{Asymptotic GW fluxes}
\label{sec:comparison_fluxes}

\begin{figure}[t]
  \hspace{-0.51cm}
  \centering  
  \includegraphics[width=0.49\textwidth]{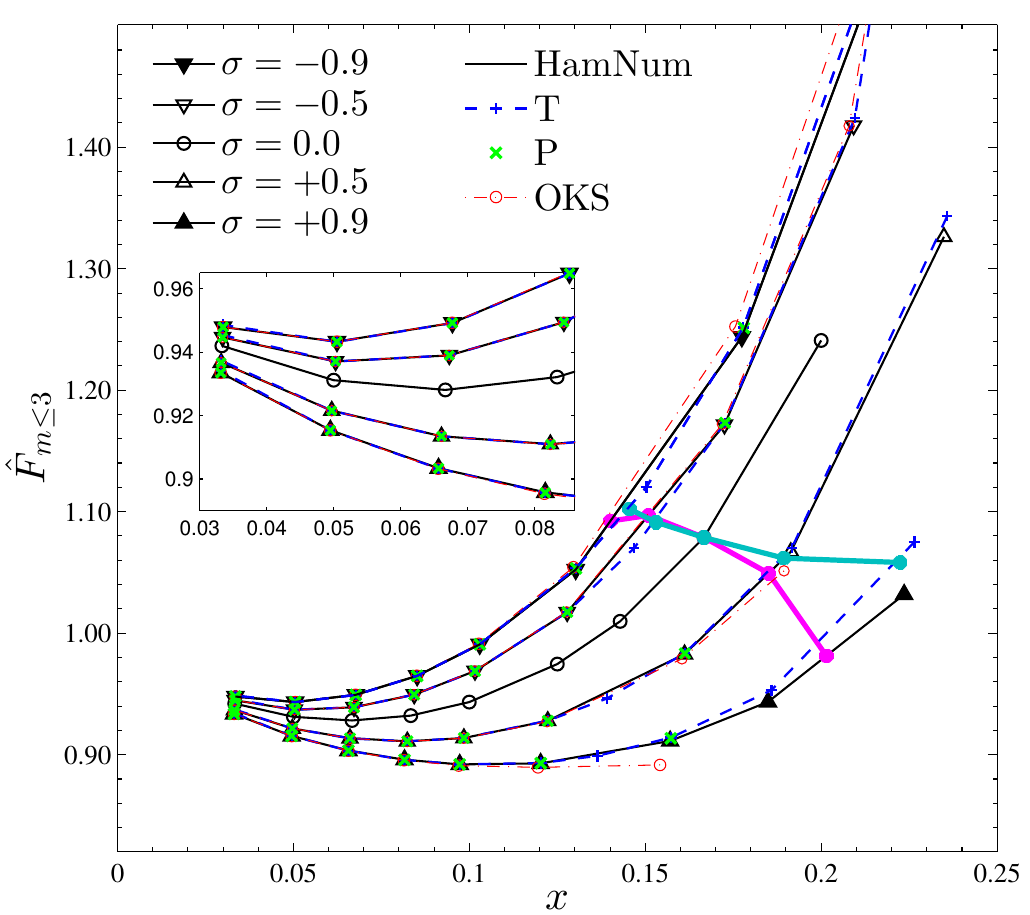} 
  \caption{%
   Comparison of the full GW energy flux, approximated by the sum of the $m=1,2,3$ modes, containing all $\ell$ contributions,
   over the variable $x=(M \Omega)^{2/3}$, where $\Omega$ is the particle's orbital frequency.
   We contrast four different circular dynamics of a spinning particle around a Schwarzschild BH;
   Hamiltonian dynamics (solid black, triangles),
   Mathisson-Papapetrou dynamics with the T SSC (blue dashed, pluses), the P SSC (green crosses),
   and the OKS SSC (red dash-dotted, circles).
   We consider the four spin values ${\sigma=-0.9,-0.5,0.5,0.9}$,
   and the nonspinning particle limit (solid black, circles).
   Additionally, the fluxes at the ISCOs are connected along the different spins
   for the Hamiltonian case (thick magenta) and the T case (thick cyan).}
  \label{fig:F_lsum_m123_over_x_differentEOMs}
\end{figure}

\begin{figure}[t]
  \centering  
  \includegraphics[width=0.44\textwidth]{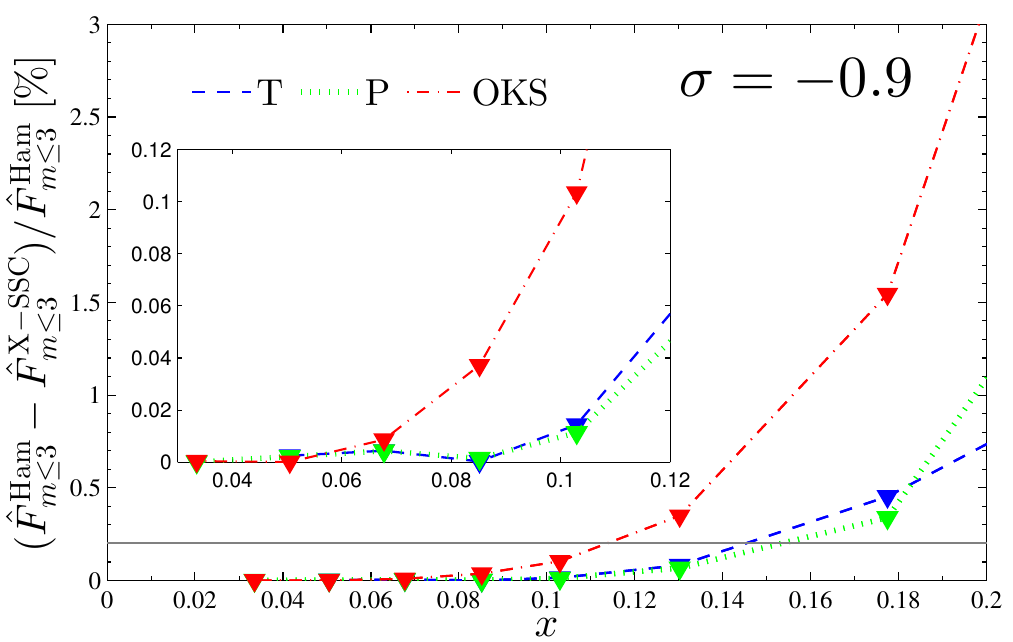}
  \includegraphics[width=0.44\textwidth]{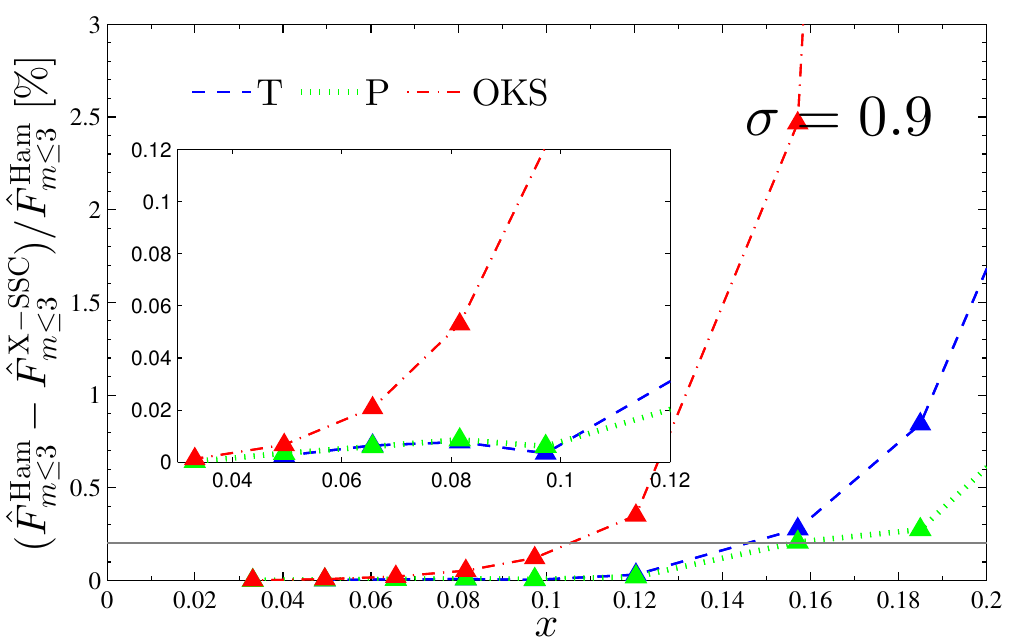}
  \caption{
  Relative differences in the full GW flux, approximated as the sum over the $m=1,2,3$ modes,
  produced using the T SSC (blue, dashed), the P SSC (green, dotted), and the OKS SSC (red, dash-dotted)
  with respect to the Hamiltonian reference case.
  We consider the two dimensionless particle spins $\sigma=-0.9$ (top panel) and $\sigma=0.9$ (bottom panel).   
  The gray horizontal lines at $0.2 \%$ mark our estimated relative numerical accuracy.
  The figures show that the GW fluxes of the various dynamics agree with one another
  at the level of our accuracy up to $x\lesssim0.11$ ($r\gtrsim 10M$).
  At smaller radii the agreement is still at the $\sim 1 \%$ level
  almost up to $r=5 M$, except for positive spins with the OKS SSC. In this sense the
  OKS SSC shows a different strong-field-behavior than the other tested SSCs.  
     }
  \label{fig:GWfluxes_over_x_reldiffs}
\end{figure}

We now compare the GW fluxes produced by the different circular dynamics
of the spinning particle discussed in
Sec's~\ref{sec:MP_dynamics}-\ref{sec:HamDynamics}. 
We briefly explain the numerical algorithm employed in the
\texttt{Teukode} for computing the waveforms; more details can be found in Paper~\RM{1}.
For an overview of the existing literature on the topic of GWs from a
spinning particle
see~\cite{Mino:1995fm,Saijo:1998mn,Han:2010tp,Tominaga:2000cs,Tominaga:1999iy,Burko:2015sqa}.

\subsection{Waveform generation algorithm}
\label{sec:teukode}

We compute the GW fluxes associated with the different circular dynamics
using the waveform generation algorithm developed in
\cite{Bernuzzi:2010ty,Bernuzzi:2011aj,Harms:2013ib,Harms:2014dqa,Harms:2015ixa}.
The latter approach is based on solving the Teukolsky equation
(TE)~\cite{Teukolsky:1972my,Teukolsky:1973ha} in (2+1)D form
on hyperboloidal slices of the Kerr spacetime~\cite{Calabrese:2005rs,Zenginoglu:2007jw,Zenginoglu:2009hd,Zenginoglu:2010cq,Vano-Vinuales:2014koa},
and including the pole-dipole particle source term. Here we briefly summarize the main
features of the method, referring to our previous work for details on numerical convergence
and thorough cross-checks with literature results in the nonspinning test-particle limit~\cite{Bernuzzi:2012ku,Bernuzzi:2010xj,
Barausse:2011kb,Sundararajan:2010sr,Sundararajan:2007jg,Hughes:1999bq,Taracchini:2014zpa}.

The TE is formulated using horizon-penetrating and
hyperboloidal coordinates following Zenginoglu's scri-fixing approach
\cite{Zenginoglu:2007jw,Zenginoglu:2009hd,Zenginoglu:2010cq}. This 
technique allows us to measure the GW signal at
future null-infinity (scri), where it is unambiguously defined.
The used approach is advantageous for numerical treatment because 
(i) the horizon and scri are included in the computational domain,
(ii) outgoing (ingoing) radial coordinate light speeds vanish at the horizon
(at scri), so no boundary conditions are 
needed. The particular coordinates employed here are the HH$_{10}$ coordinates
introduced in~\cite{Harms:2014dqa}, see also~\cite{Yang:2013uba}. The (3+1)D TE in these coordinates
is then decomposed exploiting the axisymmetry of the
background by separating each Fourier $m$-mode 
in the azimuthal direction. This results in (2+1)D wave-like equations for
each $m$-mode of the radius rescaled Weyl scalar, $r\Psi_4$. From the Weyl scalar
we reconstruct the multipoles $h_{m}$ of the metric
waveform and then decompose it into spin-weighted spherical harmonics $h_{\ell m}$~\cite{Harms:2014dqa}.

Numerical solutions of the TE in the time-domain are obtained using
the \texttt{Teukode}: a computer code specifically designed for particle
perturbations of a rotating BH~\cite{Harms:2013ib,Harms:2014dqa,Nagar:2014kha,Harms:2015ixa}. 
The TE is written as a first-order in time, and second-order in space
system  and discretized in time using the
method-of-lines. The spatial 2D domain is represented by a uniform mesh
${(y,\theta)\in[y_\text{horizon},y_\textrm{scri}]\times(0,\pi)}$, where
$y$ is the radial HH$_{10}$ coordinate, with $N_y \times N_\theta$ grid points.
Finite differencing approximations are used for the
spatial derivatives. In this work we have used $6$th order finite-differencing stencils and
employed a resolution of $N_y \times N_\theta=4800\times400$.

To assess the accuracy of our numerical results, we compared in Paper~\RM{1}
our fluxes for the nonspinning particle case against the highly accurate
reference solutions of S.~Hughes, which were computed using an improved version of
the frequency-domain code appearing in~\cite{Hughes:1999bq,Hughes:2001jr,Sundararajan:2007jg}.
Assuming that the spin of the particle does not significantly deteriorate the numerical accuracy,
we estimated our relative numerical accuracy level at $\sim 0.2 -0.5 \%$,
depending on whether the full flux or a dominant multipole, or a subdominant multipole, is considered.
The same accuracy estimate holds for the present study.
We mention that the results for the T SSC were already published in Paper~\RM{1},
where we had employed a higher resolution of $6000\times500$ points
for the outermost orbits at $r=30M$ in order to have increased accuracy in the weak-field.
In the current study we have, however, used $4800\times400$ points allover.
So the $r=30M$ T SSC results are a bit more accurate than the other
results presented here.

\subsection{Results}
\label{sec:results}

We computed the fluxes for each of the four dynamics
at various BL-radii, usually $r/M \in \{4,5,6,8,10,12,20,30\}$,
and for the four particle spins~$\sigma=\pm0.5,\pm0.9$.
The multipolar GW fluxes read
\begin{align}
\label{eq:Flux}
F &= \sum_{m=1}^{\infty} F_{m}\\
  &= \sum_{\ell=2}^{\infty}\sum_{m=1}^{m=\ell} F_{\ell m} \nonumber 
  = \dfrac{2}{16\pi}\sum_{\ell=2}^{\infty}\sum_{m=1}^{m=\ell}(m\Omega)^2| r h_{\ell m}|^2 ,
\end{align}
where we follow the notation of~\cite{Damour:2008gu}. Note that $F_m$ and $F_{\ell m}$ are defined
to contain both the $m$ and $-m$ contributions, which are equivalent for GWs from a particle on CEOs.
GW fluxes are typically studied as functions of the frequency
parameter $x$, see Eq~\eqref{eq:def_x}.
Furthermore, it is convenient to use the normalization
\begin{align}
 \label{eq:LO_normalised_fluxes}
\hat{F}_{\ell m}(x,\sigma)  & = \frac {F_{\ell m}(x,\sigma)} {  F^{\rm LO}_{\ell m}(x) } \, \, ,
\end{align}
where $ F^{\rm LO}_{\ell m}(x) $ is the leading-order (LO) flux predicted
by the quadrupole formula. Our waveform algorithm directly provides the fluxes
$F_m$ with all $\ell$-contributions included; $F_{\ell m}$ is computed
from the projected metric multipoles $h_{\ell m}$. We further define
\be
\hat{F}_{m} = \begin{cases}
\frac{F_m}{F^{\rm LO}_{2m}} & m=1 \vspace{4 pt}\\ 
\frac{F_m}{F^{\rm LO}_{mm}} & \textrm{otherwise} \ 
\end{cases} 
\ee
in order to present LO-normalized $m$-mode fluxes.

For each dynamics we have computed the fluxes for the modes ${m=1,2,3}$. The
$\hat{F}_m$ are reported in Tab~\ref{Tab:comparison_fluxes_Fm2Fm1} 
and Tab~\ref{Tab:comparison_fluxes_Fm3} for $m=1,2$ and $m=3$ respectively.
The results are illustrated in
Fig~\ref{fig:F_lsum_m123_over_x_differentEOMs}, which reports the total
normalized fluxes, i.e.\ the \textit{sum} of the $\hat{F}_m$
(see Fig~\ref{fig:F_lsum_m_over_x_differentEOMs} for the separate $m$-mode fluxes).
The fluxes from the Hamiltonian dynamics are shown in solid black,
with up- and downward triangles distinguishing the different $\sigma$.  
The fluxes from the T SSC are shown as small blue pluses, connected by a thin, dashed line.
The fluxes from the P SSC are not connected by lines, but just presented by green crosses.
Finally, the fluxes for the OKS SSC are given by red dash-dotted lines with empty circles.
For the Hamiltonian case and for the T SSC, we have additionally
included lines that connect the interpolated fluxes  
at the ISCOs for the five different spins (magenta for Hamiltonian ISCOs and cyan for T SSC ISCOs).
These ISCO lines are omitted for the OKS and the P SSC. 
The data for the nonspinning case, shown as black open circles connected by a solid line,
were computed by S.~Hughes in the frequency-domain at very high accuracy and kindly made
available to us, see~\cite{Hughes:1999bq,Hughes:2001jr,Sundararajan:2007jg,Taracchini:2013wfa}.
The differences between the various fluxes with respect to the fluxes
from the Hamiltonian dynamics are shown in
Fig~\ref{fig:GWfluxes_over_x_reldiffs} for ${\sigma=\pm0.9}$.

The figures show that for all $\sigma$ values and up to
${x\lesssim0.12-0.16}$ ($r\gtrsim 10-8M$) all the fluxes agree with the
Hamiltonian case within our numerical accuracy (${\sim0.2 \%}$). Notably, for
$x\lesssim0.04$ ($r\gtrsim20 M$) 
the relative differences are at a level of $\sim 0.02 \%$ in all cases tested. 
For the negative spins $\sigma=-0.9$ and $\sigma=-0.5$, the flux
differences remain at the $\sim 1\%$ level even beyond the T-SSC-ISCOs
at $x_{\rm{ISCO}}\approx 0.146$ and $x_{\rm{ISCO}}\approx 0.153$ respectively.
For positive spins $\sigma=+0.5$ and $\sigma=+0.9$, the agreement is slightly worse
and especially the OKS case exhibits
a systematic deviation for $x\gtrsim0.12$.
We note that our assumption on the coordinate transformation from
the EOB to the BL radial coordinate might contribute to differences seen
in the fluxes\eqref{eq:UEOB_to_uBL_linearinsigma}
(because the BL r enters the source term of the TE).

In conclusion, the differences in the fluxes for $x\gtrsim0.1$ are consistent
with our analysis of the dynamics' energetics in Sec~\ref{sec:dynamics_energeticsANDisco}.
The present analysis of the fluxes constitutes a further means to
probe that the different dynamical prescriptions are not completely equivalent
at large frequencies and large spins.
Additionally, the computation of the fluxes from the Hamiltonian dynamics
paves the way for applying a resummation/factorization procedure
along the lines of~\cite{Damour:2008gu} to Post-Newtonian
analytical test-particle waveforms that account for the spin.

Finally, we mention again that the $r=30M$ T SSC results were obtained
at higher resolutions than all other cases, which is why the relative differences in
Tabs~\ref{Tab:comparison_fluxes_Fm2Fm1}-\ref{Tab:comparison_fluxes_Fm3}
seem larger and are thus shown in brackets.

\section{Conclusions}
\label{sec:conclusions}

In this paper we have studied, for the first time, four different approaches for 
the dynamics of a spinning test-body in circular equatorial orbits around a BH, namely
i)~MP with the T SSC,
ii)~MP with the P SSC,
iii)~MP with the OKS SSC, and
iv)~Hamiltonian dynamics based on the EOB-Hamiltonian of Damour~$\&$~Nagar~\cite{Damour:2014sva}.
For each case we have presented numerical procedures for finding
initial data that upon evolution lead to circular orbits.
Notably, this is the first time that CEOs have been studied
within the OKS SSC, and the used numerical procedure is novel,
as well as the one presented for the P SSC.

The analysis of energetics and the ISCO shift (coordinate invariant
quantities) indicated that all prescriptions are practically equivalent at small
orbital frequencies $x \ll x_\text{ISCO}$. This result is expected, but
a numerical proof, and especially an analysis in the strong-field
regime, were missing in the literature. 
For $\sigma \lesssim 0.2$, all different
approaches remain very compatible up to the ISCO for negative
spins (see Fig~\ref{fig:Ex}). We also find agreement for large
negative particle spins. For $\sigma > 0.2$, however, the dynamics
exhibit drastic deviations. 
We find that, in this regime of large positive spins, 
the OKS SSC shows the most deviations from the Hamiltonian reference case;
e.g.\ the spin-orbit contribution
of the OKS SSC case and the Hamiltonian case differ by $10\%$ for $|\sigma|=0.9$ at $x \approx 0.17$. 

We have also explored the influence of the different spinning test
body dynamics on the GW fluxes. 
This analysis provides an important tool to assess
if the different dynamics are, in practice, equivalent.
Consistently with the analysis of the energetics, we found 
that the GW fluxes of the different CEO dynamics
are equivalent up to the $ 1\%$ level even at moderately large
particle frequencies $x \lesssim 0.15$ ($r \gtrsim 8 M$).
At larger distances the agreement is below our numerical uncertainty ($0.2\%$),
which gives confidence that indeed all prescriptions are compatible in the weak-field.
On the contrary, the disagreement at small $x$ indicates that at the pole-dipole
level the choice of SSC influences relevant features of the described physics.

The main practical application of this work lies 
in the context of modeling GW fluxes, as done, e.g., in the EOB model.
The EOB model relies on an analytic radiation reaction force that accounts for the GW fluxes.
This radiation reaction is found through elaborate resummation and factorization procedures~\cite{Damour:2008gu,Nagar:2016ayt},
which are currently not incorporating the spin of the particle.
The present study constitutes the first step towards including the spin
by providing a numerical target solution that guides the resummation and is needed to assure success.
While on the analytical side the literature holds available all necessary
Post-Newtonian results on the multipolar waveforms~\cite{Tanaka:1996ht,Blanchet:2011zv,Blanchet:2012sm,Marsat:2013wwa,Bohe:2015ana},
on the numerical side only the results for the T SSC dynamics existed~\cite{Harms:2015ixa}.
Before the present study it was, however, unclear to what extent the T SSC dynamics would be compatible with
the Hamiltonian dynamics that are related to the EOB approach.
Therefore, the found equivalence of the various dynamics at large orbital distances
is important since it means that a representation which is good for one case,
say the Hamiltonian dynamics, is also good for the other cases.
However, close to the ISCO it will be essential in modeling the GW fluxes
to actually use the fluxes obtained within the Hamiltonian dynamics.

A generalization of our results to the setup of a spinning test body orbiting
around a rotating BH  
has been theoretically prepared in this work and will be explored in practice 
in a subsequent work. Furthermore, it would be interesting to
check additional SSCs such as the Newton-Wigner SSC, and to include explicitly
the canonical spinning particle Hamiltonian of~\cite{Barausse:2009aa,Vines:2016unv} to the comparison.
It is also conceivable to focus on another class of orbits, e.g.\ radial infalls. As for CEOs,
one would be able to compare the different SSCs for the same physical situation,
and thus to get further grasp on the implications and potential pathologies of the different approaches.

The data computed in this work, including multipolar fluxes and the key numbers of the dynamics as presented
in Tabs~\ref{Tab:comparison_EOM_Omega&vut}-\ref{Tab:comparison_fluxes_Fm3}, are freely available at \cite{harms_2016_61308}.

\begin{acknowledgments}
We want to thank  D.~Bini, F.~Costa, T.~Damour, D.~Hilditch, B.~Br{\"u}gmann, G.~Sch{\"a}fer,
and M.~\v{S}r\'{a}mek for useful discussions
on the topic and helpful comments on the manuscript.
Special thanks to O.~Semer\'{a}k for very enlightening discussions about MP
and SSCs. 
We are particularly grateful to Scott Hughes for providing his data
on circular orbits for a nonspinning particle.
This work was supported in part by  DFG grant SFB/Transregio~7
``Gravitational Wave Astronomy''. 
E.H. thanks IHES for hospitality during the development of part of this work.
G.L-G is supported by UNCE-204020 and by GACR-14-10625S.
\end{acknowledgments}

\appendix

\section{ISCOs for a nonspinning particle}
\label{sec:ISCO_nonspinning}

The ISCO for a nonspinning particle is naturally found when writing
the geodesic EOM in the radial direction in the form
\begin{align}
  \label{eq:classical_harmonic_oscillator}
  \frac{1}{2} \left( \frac{ d r}{d \lambda} \right)^2 + V_{\rm{eff}}(J_z,r) = \epsilon \; ,
\end{align}
where $V_{\rm{eff}}(J_z,r)$ is called the effective potential of radial motion
and $\epsilon=\frac{1}{2}E^2$, with $E$ being the conserved energy of the particle
and $J_z$ the conserved z-component of its angular momentum.
The form of Eq~\eqref{eq:classical_harmonic_oscillator}
is chosen such to resemble the classical EOM of a harmonic oscillator with unit mass.
For $V_{\rm{eff}}(J_z,r)=\epsilon$ one gets turning points of radial motion.
For circular orbits the radial acceleration needs to vanish, which 
is the case at extrema of the radial potential, i.e.\ when 
\begin{align}
 \label{eq:def_circular_orbits}
 \frac{d}{dr} V_{\rm{eff}}(J_z,r) = 0 \; .
\end{align}
The circular orbit is stable/unstable when the extremum is a minimum/maximum.
If we  assume that the energy has the specific value corresponding to an ISCO,
then for a given, sufficiently large value of $J_z$ there are always two solutions
to Eq~\eqref{eq:def_circular_orbits}; one radius at which the given $J_z$ leads
to a stable circular orbit and another one that leads to a unstable circular orbit.
As $J_z$ is decreased, these two solutions approach one another until 
they coincide. For smaller $J_z$ circular orbits are ruled out because
the particle has too little angular motion to prevent itself from falling into the BH.
The solution of Eq~\eqref{eq:def_circular_orbits} which is associated to the minimum possible
value for $J_z$ is typically understood as the ISCO. We note that this orbit
corresponds to an inflection point of $V_{\rm{eff}}$, i.e.\
$\frac{d^2}{dr^2} V_{\rm{eff}} = 0$. Thus, the orbit is neither stable nor
unstable in the sense explained above, and the term ``innermost stable'' circular orbit is actually a bit misleading.
Therefore, we prefer to call that radius the ``indifferently stable circular orbit''.

To get an impression let us consider some values.
For a background spin of $\hat{a} =-0.9$
one finds $r_{\rm{ISCO}}\approx8.7$,
for $\hat{a} =0$ at $r_{\rm{ISCO}}=6M$,
and for $\hat{a} =0.9$ at $r_{\rm{ISCO}}\approx 2.3M$. 
Thus, it depends on the background spin whether an orbit at, say, $r=8M$
is a rather ``strong-field'' or ``weak-field'' orbit.

\section{Solution methods for CEOs and ISCOs}
\label{sec:solution_methods}
We want to briefly explain the specific numerical solution methods that we
employ for solving the equations that define CEOs and ISCOs,
as derived in Sec~\ref{sec:MP_dynamics}.

In general we find it convenient to use the dimensionless particle spin
$\sigma=S/(\mu M)$. If needed, one can rescale accordingly later. This means,
however, that in the calculation we have to make all the other quantities
dimensionless as well, e.g., 
 \begin{align}
  \label{eq:dimless_quantities}
  \hat{r}=\frac{r}{M},  \quad \hat{J}_z=\frac{J_z}{\mu M}, \quad
  \hat{E}=\frac{E}{\mu} \ .  
 \end{align}
The dimensionless quantities are equivalent to the dimensional quantities when
one sets $\mu=M=1$, as we do in practice. 
For more details see Sec~\RM{2}C in Paper~\RM{1}.
Note that for the P~SSC one has to use the mass $\textsf{m}$ instead of the mass
$\mu$, i.e.\ $\sigma=S/(\textsf{m} M)$, since $\textsf{m}$ is a constant of motion
while $\mu$ is not (see Tab~\ref{tab:SSC_constants}). Therefore, for the
P~SSC we set $\textsf{m}=M=1$ in our calculations. 

To find CEOs for the MP with the T SSC, we solve the system~\eqref{eq:VeffCEO_T} for a 
given radial distance $r$ and spin $S$ to get the energy $E$, and the z-component
of the total angular momentum, $J_z$. For simplicity, we solve the system using
the routine \verb#Solve# of \verb#Mathematica#. The latter gives the
same result, up to machine precision, as the closed-form solution of
\cite{Hackmann:2014tga,Tanaka:1996ht}. To find an ISCO 
for a given spin $S$, we employ
Eqs~\eqref{eq:VeffCEO_T} along with the condition 
$\frac{d^2 V_\textrm{eff,T}}{d r^2} = 0$ and also use the \verb#Solve# routine.

To find initial conditions for CEOs under the P and the OKS~SSCs,
we have to solve the respective three potential systems described in Sec~\ref{sec:CEOs_SSC_P}-\ref{sec:CEOs_SSC_OKS}.
For the P SSC this is concretely the system~\eqref{eq:PCEOs};
for the OKS~SSC the three potentials are given in Eqs~\eqref{eq:Veff_OKS},
and their derivatives with respect to $r$ can be straightforwardly computed.
In practice, to solve these systems we are using a Newton-Raphson method as
implemented in \verb#FindRoot# of \verb#Mathematica#. We used the Newton-Raphson
method, because the three potential systems are composed of equations which
contain up to sixth order polynomials in $r$ and up to second order polynomials
in the rest of the unknowns. Alternatively for the P SSC one can use \verb#NSolve#
to find CEOs; however, note that for the OKS SSC the \verb#NSolve#  misbehaves for
large radii. To verify the ability of the novel three-potentials-method to find
CEOs (Secs.~\ref{sec:CEOs_SSC_P},~\ref{sec:CEOs_SSC_OKS}), several initial
conditions have been evolved in time by the algorithm implemented already in
\cite{Lukes-Gerakopoulos:2014dma} and Paper \RM{1}.
The orbits have been found to be circular up to numerical accuracy.

Concerning the location of the ISCOs, we have used three different approaches
to test the ability of the three-potentials-method in finding them, and
these approaches give results which agree up to numerical accuracy. 
Namely, \\
a) First, we simply employed \verb#FindRoot# of \verb#Mathematica#
for a given $\sigma$ to solve the system \eqref{eq:PCEOs}
along with \eqref{eq:PISCOs} for the P SSC, and the equivalent system for the OKS SSC.
Since the numerical method for solving does not provide a unique solution,
we have chosen those solutions which appeared to be the nearest to the
analogue solution found for the T SSC for the given $\sigma$
(Tab~\ref{tab:ISCOs_x} shows these solutions). \\
b) For the second approach, we used the $a=\sigma=0$ known analytically
solution as an initial guess for the \verb#FindRoot# when $|\sigma|$ had a very
small non-zero value to calculate the new corresponding ISCO solution of the three
potential system. This new solution was then fed to \verb#FindRoot# as an
initial guess to find a solution for a little bit larger value of $|\sigma|$, 
and so on until $|\sigma|\approx 1$ has been reached
(see, e.g., Fig~\ref{fig:xISCO_shift}).\\
c) Finally, we used the CEOs' $E(J_z)$ plots for a given $\sigma$ to see where a
cusp appears. A cusp appears when the ISCO is reached and since the $E(J_z)$
plots depend on radius one can find the ISCO radius. \\

Note that in general the potential systems provide more than one solution
for CEOs. We have chosen to work with the solutions for which $1\gtrsim E>0$, $J_z>0$, 
$0<V_t\sim -E$, $0<V_\phi \sim J_z$. This applies to all three SSCs tested, so
that $V_\mu$ here refers to the reference vector in general, and it has to be
adopted appropriately according to the SSC.

\section{Analysis of dynamical data}
\label{sec:CEO_quantities}

In order to assure ourselves that our CEO initial data routines,
cf.~Sec~\ref{sec:MP_dynamics_CEOs} and Sec~\ref{sec:HamDyn_RadiusRelations}, are indeed correct,
we have integrated the respective EOM numerically as explained
in~\cite{Lukes-Gerakopoulos:2014dma}.

Having produced CEOs for all the different dynamical approaches,
inspection of the data reveals that most of the variables~\eqref{eq:spinning_particle_vars} are vanishing.
In particular, for CEOs all time derivatives are zero,
with $\frac{d}{dt} \phi \equiv \Omega$ being the only exception.
From the list~\eqref{eq:spinning_particle_vars} only the four quantities
\begin{align}
 \{ v^t, v^\phi, p_t, p_\phi \} 
\end{align}
are non-trivial. The other components of $v^\mu$ and $p_\mu$ are zero,
and the spin-tensor can be computed from the spin-parameter $\sigma$ using Eqs~\eqref{eq:SpinTensEQ}.
Note that one can compute $\Omega$ from $v^t$ and $v^\phi$.

To summarize our dynamics, Tab~\ref{Tab:comparison_EOM_Omega&vut} lists
our results for $\Omega$ and $v^t$, for the four different dynamical approaches
and for all configurations of $(r,\sigma)$ tested;
Tab~\ref{Tab:comparison_EOM_put&puphi} analogously lists the values for $p^t,p^\phi$.
Notably, these tables are enabling future studies to check our flux computations of Sec~\ref{sec:comparison_fluxes} without
the need to recompute the dynamics.
Let us look at Tabs~\ref{Tab:comparison_EOM_Omega&vut},~\ref{Tab:comparison_EOM_put&puphi}
to analyse our results.
The main observation is that the different dynamical approaches rapidly converge to the same solution
for $\{ \Omega, v^t, p_t, p_\phi \}$ as the radius increases.
For example, looking at $\Omega$ at $r=30M$, the four different approaches are equivalent at least 
in the first three significant digits. The same holds for $v^t, p_t, p_\phi$.
The equivalence is still surprisingly good at medium distances;
at $r=10M$, for instance, $v^t$ is varying at most by $\sim 0.05 \%$ between the four prescriptions.
At even smaller distances the quantities slowly start to diverge from one another.
In general one observes that the T and the P SSC yield values which are closest together,
with the Hamiltonian approach still being quite compatible; the OKS SSC values deviate the most.
Looking at Eq~\eqref{eq:v_p_TUL}, the similarity of the T SSC and the P SSC can be expected at large radii,
where the curvature is small, but it is remarkable
how much it holds at rather small radii.
After all, we conclude that the four different approaches converge to a unique result
for CEOs at large orbital distances. Thus it is clear that also the GW fluxes
have to be the same for CEOs in the weak-field.

\begin{figure*}[t]
  \centering  
  \includegraphics[width=0.31\textwidth]{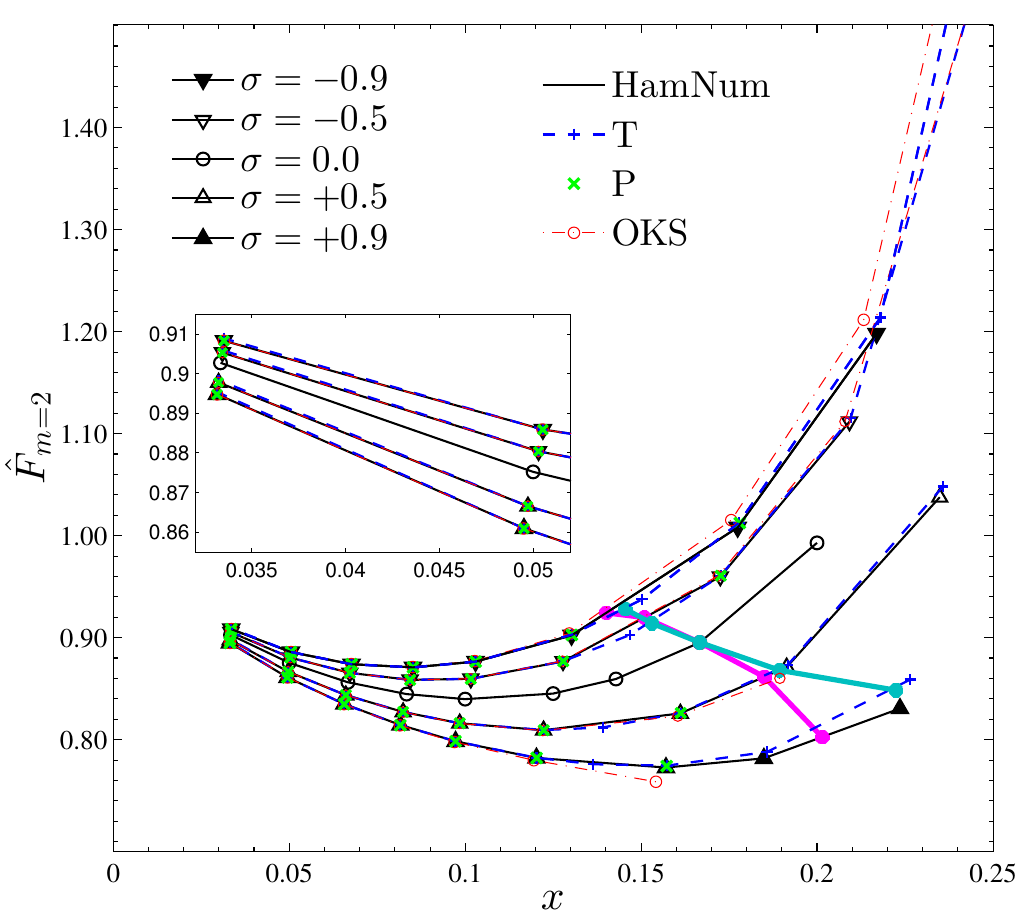}
  \includegraphics[width=0.31\textwidth]{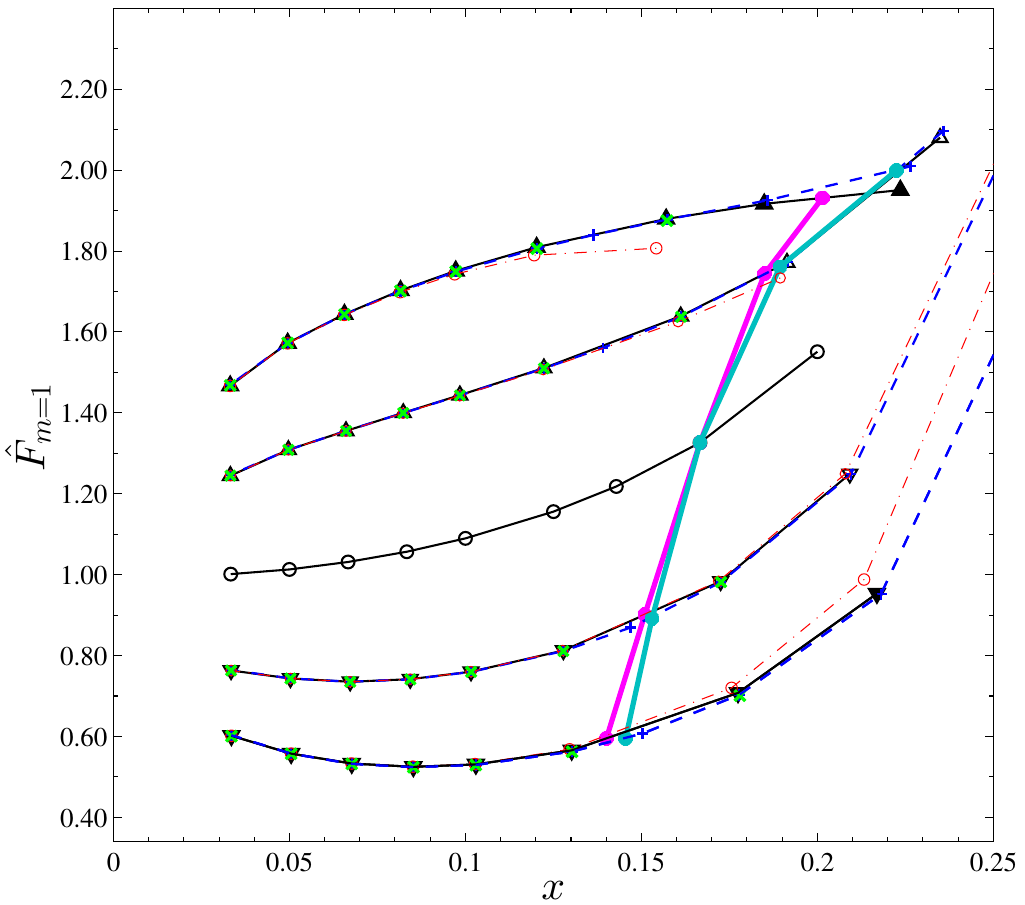} 
  \includegraphics[width=0.31\textwidth]{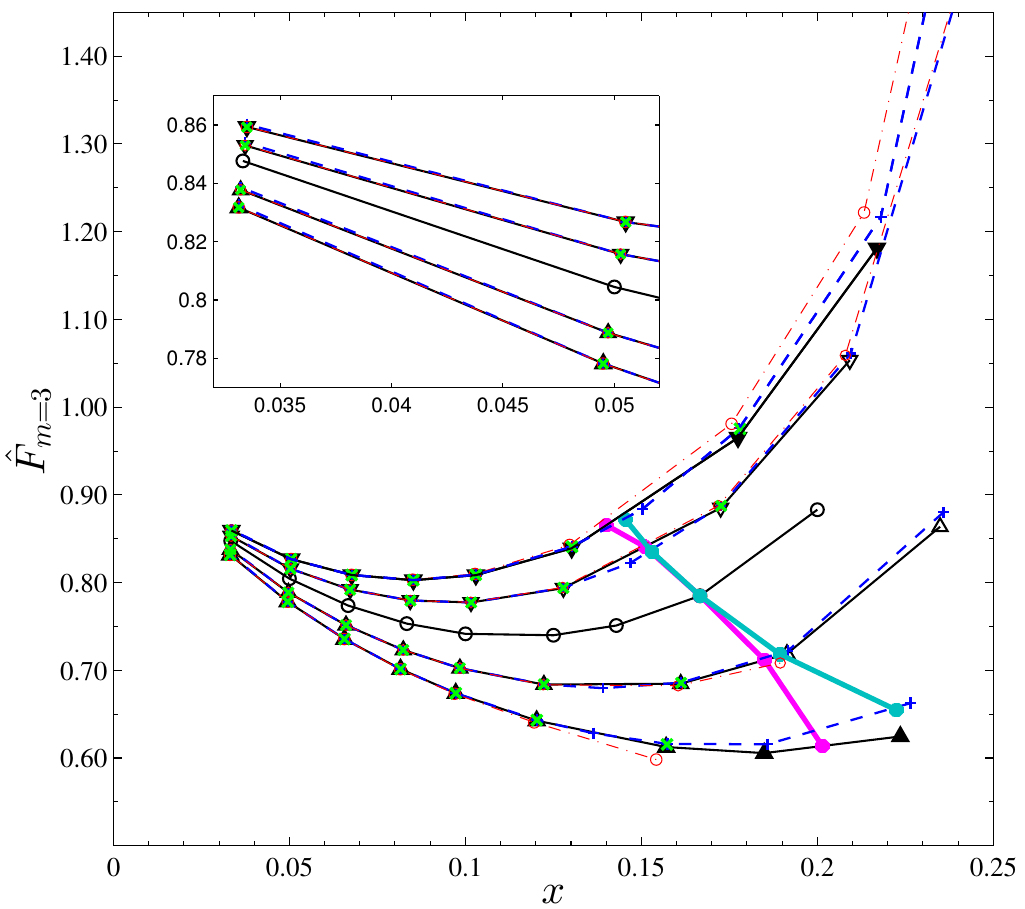} 
  \caption{
  Comparison of the multipolar $m=2$ (left), $m=1$ (center), and $m=3$ (right)
  GW fluxes over $x$, see Fig~\ref{fig:F_lsum_m123_over_x_differentEOMs} for details.
     }
  \label{fig:F_lsum_m_over_x_differentEOMs}
\end{figure*}

\begin{table*}
 
\caption{ Comparison of the dynamical quantities $\Omega$ and $v^t$ for 
          circular, equatorial orbits of a spinning particle
          around a Schwarzschild BH for
          four different prescriptions of the dynamics:
          i)~Hamiltonian dynamics (Ham),
          ii)~MP with the T SSC,
          iii)~MP with the P SSC,
          iv)~MP with the OKS SSC.
          The different cases are indicated 
          as subscripts in the respective quantities.
          The values are normalized by setting
          $\mu=M=1$.
          As the orbital distance increases all dynamics
          become equivalent. At $r=30M$ we see
          no differences in the shown five digits.
          At small radii,
          the Hamiltonian dynamics
          and the MP with the T SSC are
          still very much the same, 
          while the MP with the OKS SSC
          becomes more significantly different.
          Note that $v^\ph$ can be computed from the given
          quantities.
        }
\label{Tab:comparison_EOM_Omega&vut}

  \begin{tabular}[t]{| c | c | c c c c | c c c c |}
 \hline 
 \multicolumn{10}{|l|}{ {\bf\large{$\hat{a}=0.00$}} } \\ \hline 
 $\hat{r}$ & 
 $\sigma $ & 
 $ M \Omega_{\rm{Ham}}  $       &  $ M \Omega_{\rm{T}}  $  &  $ M \Omega_{\rm{P}}  $  &  $ M \Omega_{\rm{OKS}}  $  & 
 $ v_{\rm{Ham}}^t        $       &  $ v_{\rm{T}}^t        $  &  $ v_{\rm{T}}^t        $  &  $ v_{\rm{OKS}}^t        $    
 \\ 
 \hline 
4.00  & -0.90  & 0.14676  & 0.15052  & /  & 0.14013  & 2.53679  & 2.69700  & /  & 2.31987  \\ 
  & -0.50  & /  & 0.13805  & /  & 0.13451  & /  & 2.26404  & /  & 2.17945  \\ 
  & 0.50  & 0.11382  & 0.11452  & 0.11433  & /  & 1.84835  & 1.85641  & 1.85419  & /  \\  
  & 0.90  & 0.10573  & 0.10780  & 0.10692  & /  & 1.76463  & 1.78435  & 1.77588  & /  \\ 
  \hline 
5.00  & -0.90  & 0.10104  & 0.10184  & 0.10238  & 0.09850  & 1.70303  & 1.71313  & 1.72019  & 1.67265  \\ 
  & -0.50  & 0.09570  & 0.09592  & 0.09600  & 0.09484  & 1.64173  & 1.64407  & 1.64483  & 1.63266  \\ 
  & 0.50  & 0.08372  & 0.08390  & 0.08386  & 0.08246  & 1.53432  & 1.53570  & 1.53538  & 1.52494  \\ 
  & 0.90  & 0.07952  & 0.08009  & 0.07988  & /  & 1.50428  & 1.50821  & 1.50672  & /  \\ 
  \hline 
6.00  & -0.90  & 0.07475  & 0.07498  & 0.07512  & 0.07360  & 1.46568  & 1.46766  & 1.46878  & 1.45613  \\ 
  & -0.50  & 0.07166  & 0.07172  & 0.07174  & 0.07128  & 1.44066  & 1.44117  & 1.44132  & 1.43777  \\ 
  & 0.50  & 0.06471  & 0.06477  & 0.06476  & 0.06425  & 1.39223  & 1.39261  & 1.39252  & 1.38937  \\ 
  & 0.90  & 0.06224  & 0.06244  & 0.06237  & 0.06055  & 1.37722  & 1.37840  & 1.37797  & 1.36760  \\ 
  \hline 
8.00  & -0.90  & 0.04698  & 0.04702  & 0.04704  & 0.04666  & 1.28172  & 1.28193  & 1.28205  & 1.27969  \\ 
  & -0.50  & 0.04571  & 0.04572  & 0.04572  & 0.04560  & 1.27381  & 1.27387  & 1.27389  & 1.27319  \\ 
  & 0.50  & 0.04277  & 0.04279  & 0.04278  & 0.04266  & 1.25699  & 1.25705  & 1.25704  & 1.25639  \\ 
  & 0.90  & 0.04170  & 0.04174  & 0.04172  & 0.04133  & 1.25126  & 1.25145  & 1.25139  & 1.24932  \\ 
  \hline 
10.00  & -0.90  & 0.03303  & 0.03304  & 0.03305  & 0.03291  & 1.20309  & 1.20314  & 1.20316  & 1.20241  \\ 
  & -0.50  & 0.03239  & 0.03239  & 0.03239  & 0.03235  & 1.19945  & 1.19946  & 1.19947  & 1.19924  \\ 
  & 0.50  & 0.03089  & 0.03089  & 0.03089  & 0.03085  & 1.19134  & 1.19135  & 1.19135  & 1.19114  \\ 
  & 0.90  & 0.03033  & 0.03034  & 0.03034  & 0.03020  & 1.18845  & 1.18849  & 1.18848  & 1.18779  \\ 
  \hline 
12.00  & -0.90  & 0.02487  & 0.02487  & 0.02487  & 0.02481  & 1.15911  & 1.15913  & 1.15913  & 1.15882  \\ 
  & -0.50  & 0.02450  & 0.02450  & 0.02450  & 0.02448  & 1.15709  & 1.15709  & 1.15709  & 1.15700  \\ 
  & 0.50  & 0.02363  & 0.02363  & 0.02363  & 0.02361  & 1.15246  & 1.15246  & 1.15246  & 1.15237  \\ 
  & 0.90  & 0.02330  & 0.02330  & 0.02330  & 0.02325  & 1.15076  & 1.15077  & 1.15077  & 1.15047  \\ 
  \hline 
15.00  & -0.90  & 0.01762  & 0.01762  & 0.01762  & 0.01760  & 1.12029  & 1.12029  & 1.12029  & 1.12018  \\ 
  & -0.50  & 0.01744  & 0.01744  & 0.01744  & 0.01743  & 1.11926  & 1.11926  & 1.11927  & 1.11923  \\ 
  & 0.50  & 0.01699  & 0.01699  & 0.01699  & 0.01699  & 1.11686  & 1.11686  & 1.11686  & 1.11683  \\ 
  & 0.90  & 0.01682  & 0.01682  & 0.01682  & 0.01680  & 1.11595  & 1.11596  & 1.11596  & 1.11585  \\ 
  \hline 
20.00  & -0.90  & 0.01135  & 0.01135  & 0.01135  & 0.01135  & 1.08564  & 1.08564  & 1.08564  & 1.08561  \\ 
  & -0.50  & 0.01127  & 0.01128  & 0.01128  & 0.01127  & 1.08520  & 1.08520  & 1.08520  & 1.08519  \\ 
  & 0.50  & 0.01109  & 0.01109  & 0.01109  & 0.01109  & 1.08413  & 1.08413  & 1.08413  & 1.08412  \\ 
  & 0.90  & 0.01101  & 0.01101  & 0.01101  & 0.01101  & 1.08371  & 1.08371  & 1.08371  & 1.08368  \\ 
  \hline 
30.00  & -0.90  & 0.00614  & 0.00614  & 0.00614  & 0.00614  & 1.05442  & 1.05442  & 1.05442  & 1.05441  \\ 
  & -0.50  & 0.00611  & 0.00611  & 0.00611  & 0.00611  & 1.05427  & 1.05427  & 1.05427  & 1.05427  \\ 
  & 0.50  & 0.00606  & 0.00606  & 0.00606  & 0.00606  & 1.05392  & 1.05392  & 1.05392  & 1.05391  \\ 
  & 0.90  & 0.00604  & 0.00604  & 0.00604  & 0.00604  & 1.05378  & 1.05378  & 1.05378  & 1.05377  \\ 
  \hline 
 \end{tabular} 

\end{table*}

\begin{table*}
 
 \caption{Comparison of the dynamical quantities $p^t$ and $p^\phi$ for 
          circular, equatorial orbits of a spinning particle
          around a Schwarzschild BH. See caption of Tab~\ref{Tab:comparison_EOM_Omega&vut}
          for details.
          }
 \label{Tab:comparison_EOM_put&puphi}
  
 \begin{tabular}[t]{| c | c | c c c c | c c c c|}
 \hline 
 \multicolumn{10}{|l|}{ {\bf\large{$\hat{a}=0.00$}} } \\ \hline 
 $\hat{r}$ & 
 $\sigma $ & 
 $ p_{\rm{Ham}}^t        $       &  $ p_{\rm{T}}^t        $  &  $ p_{\rm{P}}^t        $ &  $ p_{\rm{OKS}}^t        $ & 
 $ p_{\rm{Ham}}^\phi    $       &  $ p_{\rm{T}}^\phi    $  &  $ p_{\rm{P}}^\phi    $ &  $ p_{\rm{OKS}}^\phi    $   
 \\ 
 \hline 
4.00  & -0.90  & 2.53679  & 2.47059  & /  & 2.31987  & 0.37230  & 0.35811  & /  & 0.32509  \\ 
  & -0.50  & /  & 2.22421  & /  & 2.17945  & /  & 0.30348  & /  & 0.29315  \\ 
  & 0.50  & 1.84835  & 1.84108  & 1.83930  & /  & 0.21039  & 0.20838  & 0.20792  & /  \\ 
  & 0.90  & 1.76463  & 1.74715  & 1.73973  & /  & 0.18658  & 0.18136  & 0.17931  & /  \\ 
  \hline 
5.00  & -0.90  & 1.70303  & 1.68888  & 1.69109  & 1.67265  & 0.17207  & 0.16869  & 0.16929  & 0.16476  \\ 
  & -0.50  & 1.64173  & 1.63802  & 1.63834  & 1.63266  & 0.15712  & 0.15619  & 0.15627  & 0.15483  \\ 
  & 0.50  & 1.53432  & 1.53192  & 1.53170  & 1.52494  & 0.12845  & 0.12776  & 0.12770  & 0.12574  \\ 
  & 0.90  & 1.50428  & 1.49786  & 1.49676  & /  & 0.11962  & 0.11767  & 0.11736  & /  \\  
  \hline 
6.00  & -0.90  & 1.46568  & 1.46060  & 1.46097  & 1.45613  & 0.10956  & 0.10830  & 0.10840  & 0.10718  \\ 
  & -0.50  & 1.44066  & 1.43926  & 1.43932  & 1.43777  & 0.10323  & 0.10287  & 0.10289  & 0.10249  \\ 
  & 0.50  & 1.39223  & 1.39120  & 1.39115  & 1.38937  & 0.09009  & 0.08980  & 0.08978  & 0.08927  \\ 
  & 0.90  & 1.37722  & 1.37434  & 1.37406  & 1.36760  & 0.08572  & 0.08485  & 0.08478  & 0.08281  \\ 
  \hline 
8.00  & -0.90  & 1.28172  & 1.28053  & 1.28057  & 1.27969  & 0.06022  & 0.05992  & 0.05993  & 0.05971  \\ 
  & -0.50  & 1.27381  & 1.27346  & 1.27347  & 1.27319  & 0.05822  & 0.05813  & 0.05813  & 0.05806  \\ 
  & 0.50  & 1.25699  & 1.25671  & 1.25670  & 1.25639  & 0.05377  & 0.05369  & 0.05369  & 0.05360  \\ 
  & 0.90  & 1.25126  & 1.25042  & 1.25039  & 1.24932  & 0.05218  & 0.05194  & 0.05193  & 0.05163  \\ 
  \hline 
10.00  & -0.90  & 1.20309  & 1.20268  & 1.20268  & 1.20241  & 0.03974  & 0.03964  & 0.03964  & 0.03958  \\ 
  & -0.50  & 1.19945  & 1.19933  & 1.19933  & 1.19924  & 0.03885  & 0.03882  & 0.03882  & 0.03880  \\ 
  & 0.50  & 1.19134  & 1.19123  & 1.19123  & 1.19114  & 0.03680  & 0.03677  & 0.03677  & 0.03675  \\ 
  & 0.90  & 1.18845  & 1.18812  & 1.18811  & 1.18779  & 0.03604  & 0.03596  & 0.03596  & 0.03587  \\ 
  \hline 
12.00  & -0.90  & 1.15911  & 1.15893  & 1.15893  & 1.15882  & 0.02882  & 0.02878  & 0.02878  & 0.02875  \\ 
  & -0.50  & 1.15709  & 1.15704  & 1.15704  & 1.15700  & 0.02835  & 0.02833  & 0.02833  & 0.02833  \\ 
  & 0.50  & 1.15246  & 1.15241  & 1.15241  & 1.15237  & 0.02723  & 0.02722  & 0.02722  & 0.02721  \\ 
  & 0.90  & 1.15076  & 1.15061  & 1.15060  & 1.15047  & 0.02681  & 0.02678  & 0.02678  & 0.02674  \\ 
  \hline 
15.00  & -0.90  & 1.12029  & 1.12022  & 1.12022  & 1.12018  & 0.01974  & 0.01973  & 0.01973  & 0.01972  \\ 
  & -0.50  & 1.11926  & 1.11924  & 1.11924  & 1.11923  & 0.01952  & 0.01951  & 0.01951  & 0.01951  \\ 
  & 0.50  & 1.11686  & 1.11684  & 1.11684  & 1.11683  & 0.01898  & 0.01898  & 0.01898  & 0.01897  \\ 
  & 0.90  & 1.11595  & 1.11589  & 1.11589  & 1.11585  & 0.01877  & 0.01876  & 0.01876  & 0.01875  \\ 
  \hline 
20.00  & -0.90  & 1.08564  & 1.08562  & 1.08562  & 1.08561  & 0.01232  & 0.01232  & 0.01232  & 0.01232  \\ 
  & -0.50  & 1.08520  & 1.08519  & 1.08519  & 1.08519  & 0.01224  & 0.01223  & 0.01223  & 0.01223  \\ 
  & 0.50  & 1.08413  & 1.08412  & 1.08412  & 1.08412  & 0.01202  & 0.01202  & 0.01202  & 0.01202  \\ 
  & 0.90  & 1.08371  & 1.08370  & 1.08370  & 1.08368  & 0.01194  & 0.01193  & 0.01193  & 0.01193  \\ 
  \hline 
30.00  & -0.90  & 1.05442  & 1.05441  & 1.05441  & 1.05441  & 0.00647  & 0.00647  & 0.00647  & 0.00647  \\ 
  & -0.50  & 1.05427  & 1.05427  & 1.05427  & 1.05427  & 0.00645  & 0.00645  & 0.00645  & 0.00645  \\ 
  & 0.50  & 1.05392  & 1.05391  & 1.05391  & 1.05391  & 0.00638  & 0.00638  & 0.00638  & 0.00638  \\ 
  & 0.90  & 1.05378  & 1.05377  & 1.05377  & 1.05377  & 0.00636  & 0.00636  & 0.00636  & 0.00636  \\ 
  \hline 
 \end{tabular}

\end{table*}

\begin{table*}
\caption{ Comparison of energy fluxes in the $m=2$ and $m=1$ modes produced by a spinning particle for
          four different circular dynamics: 
          i)~Hamiltonian dynamics (Ham),
          ii)~MP with the T SSC,
          iii)~MP with the P SSC,
          iv)~MP with the OKS SSC.
          We use the Hamiltonian case as the reference when computing
          the respective differences shown in the $\Delta[\%]$ columns.
          In case the relative differences fall below the level of $0.001 \%$
          we do just write $<0.001 \%$ to avoid citing more digits.          
          If a certain combination was not simulated we write a backslash $/$.
          The T SSC results for $r=30M$ were obtained at higher resolutions
          than all the other cases, see discussion in Sec~\ref{sec:teukode},
          which is why the relative differences are not consistent 
          and thus shown in brackets.
          The table compares the fluxes in the $m=1$ and $m=2$ modes
          at several Boyer-Lindquist radii $r$ and 
          for the four particle spins $\sigma=\pm0.9\pm0.5$. 
          The values for the energy fluxes have to be understood as normalized by
          the leading order Newtonian flux, cf.~Eq~\eqref{eq:LO_normalised_fluxes}.
          The main observation is that the relative differences between the respective fluxes
          vanish as the orbital distance grows.
          At $r=20M$ the energy fluxes from all dynamics agree
          in all measured cases up to $\lesssim 0.1 \%$ or better.
        }
\label{Tab:comparison_fluxes_Fm2Fm1}
\begin{tabular}[t]{| c | c | c c c c c c c | c c c c c c c |}
 \hline 
 \multicolumn{16}{|l|}{ {\bf\large{$\hat{a}=0.00$}} } \\ \hline 
 $\hat{r}$ & 
 $\sigma $ & 
 $\hat{F}_{m = 2}^{\rm{Ham}}$ &  $\hat{F}_{m = 2}^{\rm{T}}$   & $ \Delta[\%] $ & 
 $\hat{F}_{m = 2}^{\rm{P}}$   & $ \Delta[\%] $ & 
 $\hat{F}_{m = 2}^{\rm{OKS}}$ & $ \Delta[\%] $ & 
 $\hat{F}_{m = 1}^{\rm{Ham}}$ &  $\hat{F}_{m = 1}^{\rm{T}}$   & $ \Delta[\%] $ & 
 $\hat{F}_{m = 1}^{\rm{P}}$   & $ \Delta[\%] $ & 
 $\hat{F}_{m = 1}^{\rm{OKS}}$ & $ \Delta[\%] $ \\ 
 \hline 
4.00  & -0.90  & /  & 2.218  & /  & /  & /  & 2.054  & /  & /  & 2.153  & /  & /  & /  & 2.151  & /  \\ 
  & -0.50  & /  & 1.802  & /  & /  & /  & 1.741  & /  & /  & 2.298  & /  & /  & /  & 2.242  & /  \\ 
  & 0.50  & 1.038  & 1.048  & 1.017  & /  & /  & /  & /  & 2.080  & 2.097  & 0.808  & /  & /  & /  & /  \\  
  & 0.90  & 0.830  & 0.859  & 3.430  & /  & /  & /  & /  & 1.951  & 2.009  & 3.004  & /  & /  & /  & /  \\ 
  \hline 
5.00  & -0.90  & 1.198  & 1.214  & 1.330  & /  & /  & 1.212  & 1.132  & 0.953  & 0.951  & 0.170  & /  & /  & 0.988  & 3.646  \\ 
  & -0.50  & 1.112  & 1.115  & 0.324  & /  & /  & 1.112  & 0.013  & 1.249  & 1.248  & 0.057  & /  & /  & 1.249  & 0.047  \\ 
  & 0.50  & 0.871  & 0.873  & 0.232  & /  & /  & 0.860  & 1.242  & 1.772  & 1.772  & 0.053  & /  & /  & 1.733  & 2.147  \\ 
  & 0.90  & 0.782  & 0.788  & 0.804  & /  & /  & /  & /  & 1.917  & 1.925  & 0.441  & /  & /  & /  & /  \\  
  \hline 
6.00  & -0.90  & 1.008  & 1.013  & 0.457  & 1.012  & 0.420  & 1.015  & 0.705  & 0.708  & 0.703  & 0.699  & 0.701  & 1.011  & 0.720  & 1.721  \\ 
  & -0.50  & 0.960  & 0.961  & 0.105  & 0.961  & 0.105  & 0.961  & 0.084  & 0.983  & 0.982  & 0.176  & 0.981  & 0.190  & 0.984  & 0.028  \\ 
  & 0.50  & 0.826  & 0.826  & 0.067  & 0.826  & 0.057  & 0.823  & 0.313  & 1.639  & 1.638  & 0.071  & 1.637  & 0.093  & 1.625  & 0.836  \\ 
  & 0.90  & 0.773  & 0.774  & 0.232  & 0.774  & 0.164  & 0.759  & 1.810  & 1.879  & 1.878  & 0.074  & 1.875  & 0.230  & 1.807  & 3.846  \\ 
  \hline 
8.00  & -0.90  & 0.902  & 0.903  & 0.071  & 0.903  & 0.063  & 0.904  & 0.210  & 0.566  & 0.563  & 0.519  & 0.562  & 0.577  & 0.568  & 0.444  \\ 
  & -0.50  & 0.877  & 0.877  & 0.014  & 0.877  & 0.014  & 0.877  & 0.034  & 0.812  & 0.811  & 0.128  & 0.811  & 0.132  & 0.812  & 0.019  \\ 
  & 0.50  & 0.809  & 0.809  & 0.005  & 0.809  & 0.004  & 0.809  & 0.049  & 1.511  & 1.510  & 0.073  & 1.510  & 0.076  & 1.507  & 0.264  \\ 
  & 0.90  & 0.782  & 0.782  & 0.017  & 0.782  & 0.008  & 0.780  & 0.285  & 1.809  & 1.806  & 0.175  & 1.806  & 0.206  & 1.790  & 1.097  \\ 
  \hline 
10.00  & -0.90  & 0.876  & 0.876  & 0.010  & 0.876  & 0.008  & 0.877  & 0.076  & 0.531  & 0.529  & 0.311  & 0.529  & 0.327  & 0.532  & 0.136  \\ 
  & -0.50  & 0.860  & 0.860  & $ < $ 0.001  & 0.860  & $ < $ 0.001  & 0.860  & 0.013  & 0.760  & 0.759  & 0.078  & 0.759  & 0.078  & 0.760  & 0.022  \\ 
  & 0.50  & 0.816  & 0.816  & 0.002  & 0.816  & 0.002  & 0.816  & 0.014  & 1.444  & 1.443  & 0.048  & 1.443  & 0.048  & 1.442  & 0.121  \\ 
  & 0.90  & 0.798  & 0.798  & 0.008  & 0.798  & 0.010  & 0.798  & 0.083  & 1.751  & 1.749  & 0.128  & 1.749  & 0.137  & 1.743  & 0.485  \\ 
  \hline 
12.00  & -0.90  & 0.871  & 0.871  & 0.002  & 0.871  & 0.003  & 0.871  & 0.032  & 0.525  & 0.524  & 0.194  & 0.524  & 0.203  & 0.525  & 0.041  \\ 
  & -0.50  & 0.859  & 0.859  & 0.002  & 0.859  & 0.002  & 0.859  & 0.005  & 0.741  & 0.741  & 0.050  & 0.741  & 0.048  & 0.741  & 0.016  \\ 
  & 0.50  & 0.827  & 0.827  & 0.002  & 0.827  & 0.002  & 0.827  & 0.005  & 1.400  & 1.399  & 0.034  & 1.399  & 0.034  & 1.399  & 0.067  \\ 
  & 0.90  & 0.814  & 0.814  & 0.010  & 0.814  & 0.010  & 0.814  & 0.033  & 1.703  & 1.701  & 0.089  & 1.701  & 0.089  & 1.698  & 0.258  \\ 
  \hline 
15.00  & -0.90  & 0.874  & 0.874  & 0.005  & 0.874  & 0.005  & 0.874  & 0.010  & 0.533  & 0.532  & 0.114  & 0.532  & 0.122  & 0.533  & 0.004  \\ 
  & -0.50  & 0.865  & 0.865  & 0.001  & 0.865  & 0.002  & 0.865  & 0.002  & 0.735  & 0.735  & 0.020  & 0.735  & 0.022  & 0.735  & 0.009  \\ 
  & 0.50  & 0.844  & 0.844  & 0.002  & 0.844  & 0.002  & 0.844  & 0.002  & 1.355  & 1.355  & 0.013  & 1.355  & 0.019  & 1.355  & 0.037  \\ 
  & 0.90  & 0.835  & 0.835  & 0.007  & 0.835  & 0.006  & 0.835  & 0.011  & 1.644  & 1.643  & 0.048  & 1.643  & 0.052  & 1.642  & 0.129  \\ 
  \hline 
20.00  & -0.90  & 0.886  & 0.886  & 0.002  & 0.886  & 0.002  & 0.886  & 0.002  & 0.557  & 0.557  & 0.010  & 0.557  & 0.021  & 0.557  & 0.020  \\ 
  & -0.50  & 0.880  & 0.880  & 0.001  & 0.880  & 0.001  & 0.880  & $ < $ 0.001  & 0.743  & 0.743  & 0.017  & 0.743  & 0.020  & 0.743  & 0.012  \\ 
  & 0.50  & 0.867  & 0.867  & 0.002  & 0.867  & 0.002  & 0.867  & 0.002  & 1.309  & 1.309  & 0.005  & 1.309  & 0.008  & 1.308  & 0.017  \\ 
  & 0.90  & 0.861  & 0.861  & 0.003  & 0.861  & 0.004  & 0.861  & 0.003  & 1.572  & 1.572  & 0.013  & 1.572  & 0.001  & 1.572  & 0.043  \\ 
  \hline 
30.00  & -0.90  & 0.908  & 0.909  & (0.065)  & 0.908  & $ < $ 0.001  & 0.908  & $ < $ 0.001  & 0.602  & 0.603  & (0.230)  & 0.601  & 0.117  & 0.601  & 0.029  \\ 
  & -0.50  & 0.905  & 0.906  & (0.071)  & 0.905  & $ < $ 0.001  & 0.905  & 0.002  & 0.763  & 0.764  & (0.185)  & 0.763  & 0.011  & 0.763  & 0.003  \\ 
  & 0.50  & 0.898  & 0.898  & (0.065)  & 0.898  & 0.003  & 0.898  & 0.002  & 1.245  & 1.246  & (0.129)  & 1.244  & 0.015  & 1.244  & 0.014  \\ 
  & 0.90  & 0.895  & 0.895  & (0.065)  & 0.895  & $ < $ 0.001  & 0.895  & $ < $ 0.001  & 1.467  & 1.469  & (0.163)  & 1.468  & 0.049  & 1.466  & 0.038  \\ 
  \hline 
 \end{tabular}

\end{table*}

\begin{table*}

 \caption{Complement of Tab~\ref{Tab:comparison_fluxes_Fm2Fm1} for the $ m=3$ mode.
          See caption of Tab~\ref{Tab:comparison_fluxes_Fm2Fm1} for details.} 
\label{Tab:comparison_fluxes_Fm3}
\begin{tabular}[t]{| c | c | c c c c c c c |}
 \hline 
 \multicolumn{9}{|l|}{ {\bf\large{$\hat{a}=0.00$}} } \\ \hline 
 $\hat{r}$ & 
 $\sigma $ & 
 $\hat{F}_{m = 3}^{\rm{Ham}}$ &  $\hat{F}_{m = 3}^{\rm{T}}$   & $ \Delta[\%] $ & 
 $\hat{F}_{m = 3}^{\rm{P}}$   & $ \Delta[\%] $ & 
 $\hat{F}_{m = 3}^{\rm{OKS}}$ & $ \Delta[\%] $ \\ 
 \hline 
4.00  & -0.90  & /  & 2.425  & /  & /  & /  & 2.249  & /  \\ 
  & -0.50  & /  & 1.840  & /  & /  & /  & 1.776  & /  \\ 
  & 0.50  & 0.864  & 0.880  & 1.892  & /  & /  & /  & /  \\   
  & 0.90  & 0.625  & 0.663  & 6.077  & /  & /  & /  & /  \\
  \hline 
5.00  & -0.90  & 1.181  & 1.217  & 3.032  & /  & /  & 1.222  & 3.462  \\ 
  & -0.50  & 1.053  & 1.061  & 0.743  & /  & /  & 1.059  & 0.525  \\ 
  & 0.50  & 0.719  & 0.722  & 0.511  & /  & /  & 0.708  & 1.460  \\  
  & 0.90  & 0.606  & 0.616  & 1.646  & /  & /  & /  & /  \\  
  \hline 
6.00  & -0.90  & 0.965  & 0.976  & 1.123  & 0.975  & 1.054  & 0.981  & 1.650  \\ 
  & -0.50  & 0.885  & 0.887  & 0.273  & 0.887  & 0.272  & 0.888  & 0.303  \\ 
  & 0.50  & 0.685  & 0.686  & 0.183  & 0.686  & 0.171  & 0.682  & 0.326  \\ 
  & 0.90  & 0.613  & 0.616  & 0.582  & 0.616  & 0.491  & 0.598  & 2.332  \\ 
  \hline 
8.00  & -0.90  & 0.839  & 0.841  & 0.228  & 0.841  & 0.215  & 0.843  & 0.459  \\ 
  & -0.50  & 0.794  & 0.794  & 0.055  & 0.794  & 0.054  & 0.794  & 0.095  \\ 
  & 0.50  & 0.684  & 0.684  & 0.035  & 0.684  & 0.034  & 0.684  & 0.035  \\ 
  & 0.90  & 0.643  & 0.643  & 0.107  & 0.643  & 0.096  & 0.641  & 0.314  \\ 
  \hline 
10.00  & -0.90  & 0.808  & 0.809  & 0.062  & 0.809  & 0.058  & 0.810  & 0.168  \\ 
  & -0.50  & 0.777  & 0.777  & 0.015  & 0.777  & 0.014  & 0.778  & 0.036  \\ 
  & 0.50  & 0.702  & 0.702  & 0.008  & 0.702  & 0.008  & 0.702  & 0.005  \\ 
  & 0.90  & 0.674  & 0.674  & 0.023  & 0.674  & 0.021  & 0.673  & 0.077  \\ 
  \hline 
12.00  & -0.90  & 0.803  & 0.803  & 0.019  & 0.803  & 0.018  & 0.803  & 0.073  \\ 
  & -0.50  & 0.779  & 0.780  & 0.004  & 0.780  & 0.004  & 0.780  & 0.016  \\ 
  & 0.50  & 0.723  & 0.723  & 0.001  & 0.723  & 0.001  & 0.723  & 0.001  \\ 
  & 0.90  & 0.701  & 0.701  & 0.003  & 0.701  & 0.003  & 0.701  & 0.026  \\ 
  \hline 
15.00  & -0.90  & 0.809  & 0.809  & 0.002  & 0.809  & 0.002  & 0.809  & 0.025  \\ 
  & -0.50  & 0.792  & 0.792  & $ < $ 0.001  & 0.792  & $ < $ 0.001  & 0.792  & 0.005  \\ 
  & 0.50  & 0.751  & 0.751  & $ < $ 0.001  & 0.751  & $ < $ 0.001  & 0.751  & $ < $ 0.001  \\ 
  & 0.90  & 0.736  & 0.736  & 0.003  & 0.736  & 0.003  & 0.736  & 0.009  \\ 
  \hline 
20.00  & -0.90  & 0.827  & 0.827  & 0.003  & 0.827  & 0.003  & 0.827  & 0.006  \\ 
  & -0.50  & 0.816  & 0.816  & $ < $ 0.001  & 0.816  & $ < $ 0.001  & 0.816  & 0.002  \\ 
  & 0.50  & 0.789  & 0.789  & $ < $ 0.001  & 0.789  & 0.001  & 0.789  & 0.002  \\ 
  & 0.90  & 0.778  & 0.778  & $ < $ 0.001  & 0.778  & 0.002  & 0.778  & 0.001  \\ 
  \hline 
30.00  & -0.90  & 0.859  & 0.860  & (0.098)  & 0.859  & 0.007  & 0.859  & $ < $ 0.001  \\ 
  & -0.50  & 0.853  & 0.854  & (0.114)  & 0.853  & 0.006  & 0.853  & 0.006  \\ 
  & 0.50  & 0.838  & 0.839  & (0.103)  & 0.838  & 0.002  & 0.838  & 0.004  \\ 
  & 0.90  & 0.832  & 0.833  & (0.101)  & 0.832  & 0.002  & 0.832  & $ < $ 0.001  \\ 
  \hline 
 \end{tabular}

\end{table*}


\bibliographystyle{unsrt}
\bibliography{../../refs/refs}

\end{document}